\documentclass[10pt,twocolumn,prl,aps,floatfix,superscriptaddress,longbibliography]{revtex4-1}

\usepackage{hyperref}

\usepackage{listings}
\usepackage{xcolor}
\usepackage{listings}
\usepackage{hyperref}
\usepackage{lineno}
\usepackage{latexsym}
\usepackage{amssymb}
\usepackage{caption}
\usepackage{float}
\usepackage{mathrsfs}
\usepackage{subfigure}
\usepackage{graphicx}
\usepackage{amsmath}
\usepackage{cleveref}
\usepackage{subfigure}
\usepackage{color}
\usepackage{lineno}
\usepackage{pdfpages} 
\usepackage{pgffor} 
\usepackage{xr} 

\makeatletter
\AtBeginDocument{\let\LS@rot\@undefined}
\makeatother

\def\supplementfilename{supplement}

\pdfximage{\supplementfilename.pdf}
\def\numbersupplementpages{\the\pdflastximagepages}

\newif\ifarXiv
\arXivtrue 
\begin{document}
	
\preprint{APS/123-QED}

\title{Deep-learned speckle pattern and its application to ghost imaging}

\author{Xiaoyu Nie}
\affiliation{%
	Texas A\&M University, College Station, Texas, 77843, USA}%
\affiliation{%
	Xi'an Jiaotong University, Xi'an, Shaanxi 710049, China}%
\author{Haotian Song}
\affiliation{%
	Xi'an Jiaotong University, Xi'an, Shaanxi 710049, China}%
\author{Wenhan Ren}
\affiliation{%
	Texas A\&M University, College Station, Texas, 77843, USA}%
\affiliation{%
	Xi'an Jiaotong University, Xi'an, Shaanxi 710049, China}%
\author{Xingchen Zhao}%
\affiliation{%
	Texas A\&M University, College Station, Texas, 77843, USA}%
\author{Zhedong Zhang}
\email{zzhan26@cityu.edu.hk}
\affiliation{Department of Physics, City University of Hong Kong, Kowloon, Hong Kong SAR}
\author{Tao Peng}%
\email{taopeng@tamu.edu}
\affiliation{%
	Texas A\&M University, College Station, Texas, 77843, USA}%
\author{Marlan O. Scully}%
\affiliation{%
	Texas A\&M University, College Station, Texas, 77843, USA}%
\affiliation{%
Baylor University, Waco, 76706, USA}%
\affiliation{%
Princeton University, Princeton, NJ 08544, USA}%
\date{\today}

\begin{abstract}
In this paper, we present a method for speckle pattern design using deep learning. The speckle patterns possess unique features after experiencing convolutions in Speckle-Net, our well-designed framework for speckle pattern generation. We then apply our method to the computational ghost imaging system. The standard deep learning-assisted ghost imaging methods use the network to recognize the reconstructed objects or imaging algorithms. In contrast, this innovative application optimizes the illuminating speckle patterns via Speckle-Net with specific sampling ratios. Our method, therefore, outperforms the other techniques for ghost imaging, particularly its ability to retrieve high-quality images with extremely low sampling ratios. It opens a new route towards non-trivial speckle generation by referring to a standard loss function on specified objectives with the modified deep neural network. It also has great potential for applications in the fields of dynamic speckle illumination microscopy, structured illumination microscopy, x-ray imaging, photo-acoustic imaging, and optical lattices.
\end{abstract}

	\maketitle
	
Typical speckle patterns are generated when light is scattered or diffused from the inhomogeneous rough media~\cite{pine1988diffusing}. The statistics of the speckles depends on the incident light field~\cite{li2020photon}. In particular, scattered laser speckle is known as the Rayleigh speckle with a negative-exponential intensity probability density function (PDF)~\cite{goodman1975statistical}.
Speckle patterns can also be produced by sources such as x-rays~\cite{zanette2014speckle}, microwaves~\cite{wang2011transport}, and Terahertz radiation~\cite{olivieri2020hyperspectral} besides visible light. The study of speckle patterns has been conducted in many scenarios such as waveguides~\cite{valley2016multimode}, fibers~\cite{redding2013all}, and nanowires~\cite{strudley2013mesoscopic}. The wide range of applications of the speckle patterns include spectroscopy~\cite{redding2013compact}, microscopy~\cite{ventalon2006dynamic,mertz2011optical,mudry2012structured}, interferometry~\cite{nakadate1985fringe}, metrology techniques~\cite{yilmaz2015speckle,pascucci2016superresolution}, and correlated disorder optical lattices in cold atoms~\cite{mcgehee2013three,delande2014mobility,fratini2015anderson,liu2021generation}. In these applications, the speckle patterns act as efficient random carriers of encoding the spatial information within the systems and later on being decoded. For example, the non-Rayleigh non-diffractive speckle pattern is urgent needs for research on localization using optical lattice, in which a ring-shape anti-symmetric phase filter is applied~\cite{liu2021generation}. Therefore, to retain well-performed and stabilized data carriers, manipulation of its inherent statistical properties is highly demanding from the perspective of efficiency, accuracy, and robustness.

Speckle pattern also plays an essential role in ghost imaging~\cite{bennink2002two,chen2009lensless}. Standard Rayleigh speckles have been used for ghost imaging for decades~\cite{valencia2005two}. Later on, the spatial light modulator (SLM) and the digital micromirror device (DMD) are used as convenient and powerful tools for speckle pattern formation~\cite{shapiro2008computational}. Various synthesized speckle patterns~\cite{bromberg2014generating,kondakci2016sub,bender2018customizing,li2021sub,nie2021noise} are generated by customizing and regulating amplitudes and phases of the electromagnetic fields or directly designing and adjusting the power spectrum of the speckle patterns. Recently, efforts have been made to generate orthonormalized~\cite{luo2018orthonormalization}, Walsh-Hadamard~\cite{wang2016fast,zhang2017hadamard,yu2019super}, and colored noise~\cite{nie2020sub} speckle patterns for sub-Nyquist sampling imaging. 
To date, the synthesized speckle patterns are typically generated from customizing the power spectrum, vortex, amplitude of either the intensity or field distribution to justify their spatial correlations. Therefore, tremendous work must be done from complicated theoretical calculations and many experimental attempts to decide the parameters discussed above. Besides, the speckle patterns used for sub-Nyquist imaging are not optimal for any specific sampling ratio (SR).

In this work, we introduce a universally applicable speckle pattern generating method based on deep learning (DL), namely Speckle-Net. We design a specific deep neural network (DNN) customized to speckle pattern generation by utilizing the convolution concept in the convolutional neural network (CNN). The kernels in CNN are used to adjust the second-order correlation of the speckle patterns. The Speckle-Net training is a pre-processing technique only based on the optical system and the loss function. During the training process, Speckle-Net continuously improves the kernel values in each training epoch until they reach the optimum values referring to the loss function. We then implement this technique on the ghost imaging system, in which the speckle patterns are optimized for any given SR. The optimized speckle pattern can then be applied to any computational ghost imaging (CGI) system, resulting in high-quality images even at extremely low SRs. Speckle-Net can also be applied to other illumination systems that require optimizing the speckle patterns by selecting suitable evaluators as the loss function in a one-time training process. 
\section{Principles of Speckle-Net}

\subsection{Correlation Modulation by Kernels}

\begin{figure}[t]
\centering
\includegraphics[width=0.7\linewidth]{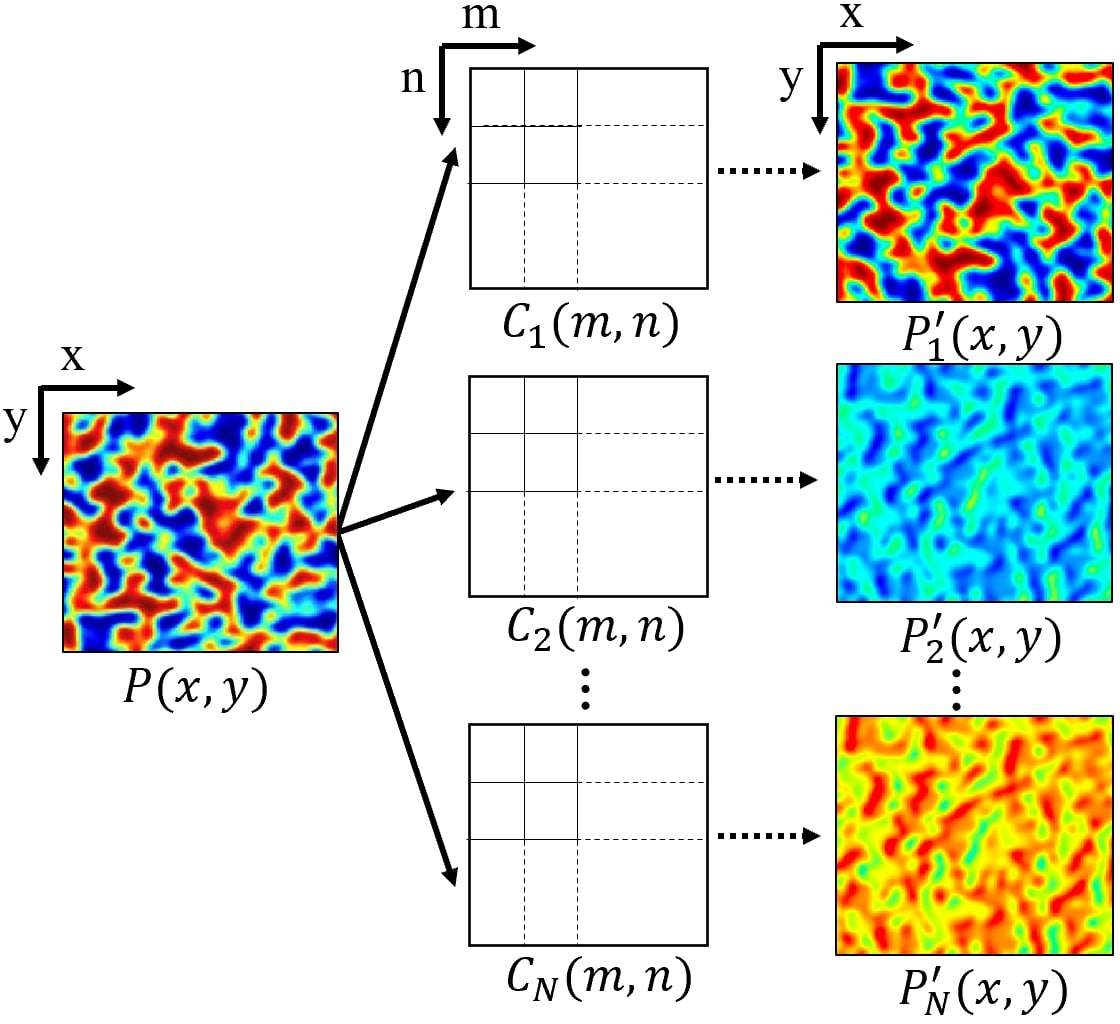}
\caption{One layer convolution in our featured neural network. Multiple kernels are attached on a single speckle pattern $P$. $P$ could be the initial pattern or convoluted pattern from the former layer, and $P'$s are the output patterns from the current layer. Each subscript indicates the correspondence between convoluted speckle patterns and kernels.}  
\label{fig:nkernels}
\end{figure}

Kernel, a popular concept in DL, is usually applied in CNN containing each convolutional layer. It functions as a matrix that makes convolution on speckle patterns to minimize the size of patterns and localizes the feature within areas in the pattern. The output patterns are modulated by convolutions between kernels and the initial pattern. In our strategy, multiple unique kernels are designed to act on initial speckle patterns $P_i(x,y)$ in each layer of DL, after which multiple different speckle patterns $P'_i(x,y)$ ($i=1,\cdots, N$) are generated, as is shown in Fig.~\ref{fig:nkernels}. Speckle patterns after multiple convolution transformations ought to be distinct from each other, and their spatial intensity fluctuation correlation distribution will be modulated by multiple kernels in multiple layers with the instruction of standard loss function in DL.

The principle of the second order correlation modulation is briefly explained here. We use in total N kernels  $C_i(m,n)$, where $m,n$ are coordinates of the kernel. The speckle patterns after the convolution can be expressed as $P'_i(x,y)=\sum_{m,n}C_i(m,n)P(x+m,y+n)$, where $x,y$ are coordinates of the patterns. The average value of the resulted patterns $P'_i(x,y)$ is
\begin{align}
\bar {P'}(x,y) & =\frac{1}{N}\sum^{N}_{i=1}\sum_{m,n}C_i(m,n)P(x+m,y+n)\cr
&=\sum_{m,n} \bar {C}(m,n) P(x+m,y+n). 
\label{eq:method1-2}
\end{align} 
We then have
\begin{align}
\Delta P'_i(x,y)&\equiv P'_i(x,y)-\bar {P'}(x,y)\cr
&=\sum_{m,n}( C_i(m,n)-\bar {C}(m,n))-P(x+m,y+n)\cr &=\sum_{m,n}\Delta C_i(m,n)P(x+m,y+n).
\label{eq:method1-4}
\end{align}
The correlation function of the resulted patterns is
\begin{align}
& \ \ \ \ \ \Gamma^{(2)}(\Delta x, \Delta y)\cr&=\langle\Delta P'_i(x_1,y_1)\Delta P'_i(x_2,y_2)\rangle\cr
&=\bigg\langle[\sum_{m_1,n_1}\Delta C_i(m_1,n_1)P(x_1+m_1,y_1+n_1)]\cr &\times[\sum_{m_2,n_2}\Delta C_i(m_2,n_2)P(x_2+m_2,y_2+n_2)]\bigg\rangle\cr
&=\sum_{m_{1,2},n_{1,2}}\langle\Delta C(m_1,n_1)\Delta C(m_2,n_2)\rangle\cr &\times P(x_1+m_1,y_1+n_1)P(x_2+m_2,y_2+n_2)\cr
&=\sum_{m_{1,2},n_{1,2}}\Gamma_C^{(2)}(\Delta m, \Delta n)\cr &\times P(x_1+m_1,y_1+n_1)P(x_2+m_2,y_2+n_2),
\label{eq:method1-3}
\end{align}
where $\Delta x\equiv x_1-x_2$, $\Delta y\equiv y_1-y_2$. It is clear shown in Eq.~(\ref{eq:method1-3}) that the correlation function of the generated speckle patterns depends on the correlation function of the kernel $\Gamma_C^{(2)}(\Delta m, \Delta n)$ and the initial pattern. Thus, the process of adjustment on each kernel in DL is aimed at producing desired correlations with respect to the initial speckle pattern, which can be seen as weight parameters. The convolution process of a single pattern can be understood as a re-distribution of the spatial correlation from different kernels. 

\subsection{Structure of the Speckle-Net}
\begin{figure*}[!ht]
\centering
\includegraphics[width=0.95\linewidth]{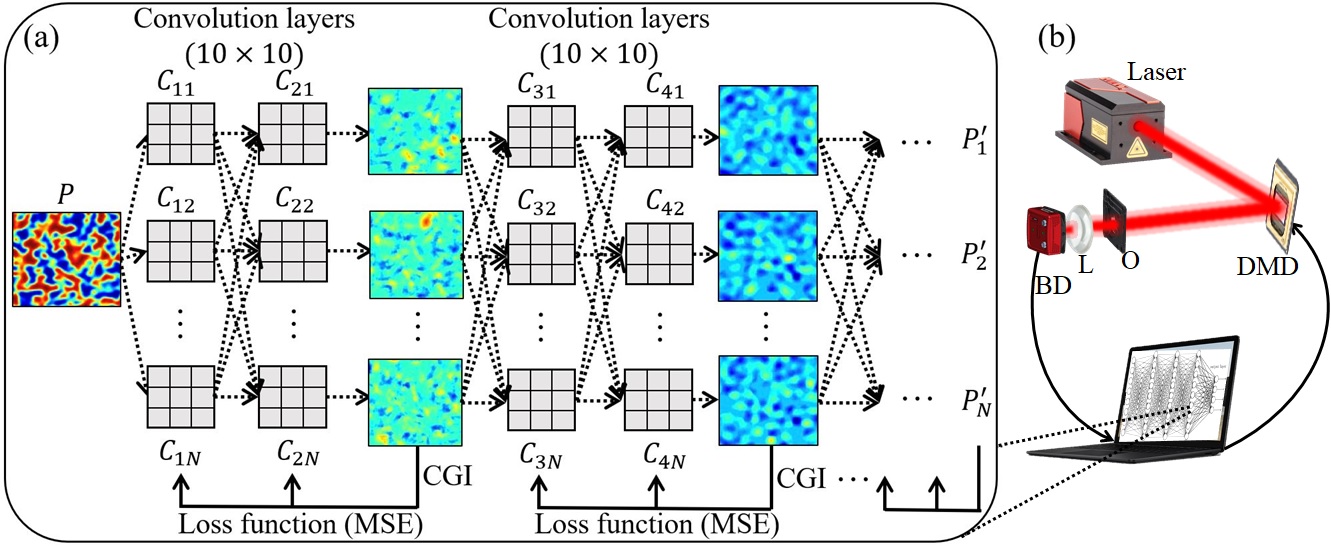}
\caption{(a) Diagram of Speckle-Net. The Speckle-Net consists of multi-branch and two convolution layers within each branch. The $10\times10$-sized kernels are adopted in each layer. The subscripts {\it j} and {\it i} in $C_{ji}$ denote the {\it j-th} layer and {\it i-th} kernel in each layer. A loss function feedback is applied at the end of each branch to modify the parameters in kernels. The deep-learned speckle patterns generate CGI results at each training epoch. (b) Schematic of the experimental setup. The deep-learned speckle patterns $P'_i$ are applied to the DMD for the CGI measurement. The laser illuminated patterns are projected onto the object (O). Light passing through object is collected by a bucket detector (BD) through a short focal length lens (L).}
\label{fig:setup_CGI}
\end{figure*}

Speckle-Net consists of multi-branch and simplified layers, as shown in Fig.~\ref{fig:setup_CGI} (a)\footnote{The raw codes of Speckle-Net can be found on \url{https://github.com/XJTU-TAMU-CGI/PatternDL}}. Single pattern padded with reflection of their boundaries plays the role of input. To provide the flexibility of correlation adjustment, convolution layers with a relatively large kernel size of $10 \times 10$, a Rectified Linear Unit (ReLU)~\cite{nair2010rectified}, and a Batch Normalization Layer (BNL)~\cite{ioffe2015batch}, are combined into a series of processes in each layer. The layers share similarities with Branch Convolutional Neural Network~\cite{BCNNpaper}, and the outputs of all layers are padded again by boundary reflections to maintain the size of their origin. The ReLU could improve the sensitivity to the activation sum input, and BNL is used to reduce internal covariate shift.

Speckle-Net offers superior efficiency and effectiveness to conventional CNN. It has no overfitting issues and can be easily adapted to other systems. Firstly, the multiple backward methods significantly improve the performance of the network. It is difficult to analyze and enhance the original pattern and aimed imaging systems from a single or a few intermediate layers. On the other hand, too many layers have poor directional of amelioration~\cite{2016ResNet}, therefore losing the characteristics of the original pattern and sought imaging systems. Nevertheless, our multi-branches neural network boosts the feedback gradient adjustment at each epoch from the loss function, avoiding the loss function of output patterns trapped in a local minimum. The parameters of the two layers in one branch are adjusted independently. Therefore, getting the optimum parameters in our model is more efficient and effective than single-branch CNN with multiple layers and single loss function feedback. Meanwhile, this Multi-branches learning process has excellent performance because various training complexities are required for different SRs. For example, when a small SR is adopted, fewer patterns lead to fewer required parameters and less time for training. Therefore, only two rounds of training are necessary to get desired speckle patterns. Otherwise, more branches can be used for a larger SR, as shown in Supplement 1. Thus, this Multi-branches Speckle-Net enables us to select the most efficient number of training branches according to the loss function values form previous results. If the loss in two (or more, to ensure) neighboring training branches goes close to the same minimum, we can conclude that the speckle patterns reach the global optimum. Secondly, we abandon the fully connected (FC) and dropout layers. FC layers in this structure demand large RAM\footnote{For instance, if the size of each speckle pattern is $112\times 112$ pixels and the SR $\beta = 0.5\%$, the number of patterns is 62. Then the size of parameters in the FC layer is around 9,000 TB, which is unrealistic for training.} and are useless in that the convolution parts aim to adjust the correlation of patterns rather than get the CGI results. On the other hand, the use of the dropout layer is to avoid over-fitting in convolutional layers. However, in a deep-learned speckle pattern scheme, the optimum patterns are our ultimate goal which remains intact for various training and testing images. A constant input image means that over-fitting does not exist in our model. Therefore, the epoch number can be determined based on the convergence of the loss function in each branch, as shown in Supplement 1. Moreover, the loss function in our model can be adjusted according to the feature of the physical process, and the CGI algorithm can be substituted by other physical processes as well. In imaging and spectroscopic systems, the mean square error (MSE), contrast-to-noise ratio (CNR), correlation-coefficient, \textit{etc.}, can be applied to the loss function independently or in combination to achieve good visibility, high contrast, and optimized similarities.

\section{Implementation: Computational ghost imaging}

Ghost imaging~\cite{pittman1995optical,bennink2002two,chen2009lensless}, a single pixel imaging technique, reconstructs the object through second-order correlation between reference and object light paths. CGI~\cite{shapiro2008computational,bromberg2009ghost} substitutes the reference path by preparing speckles in advance. Therefore, one only needs to record the intensity of object light path and correlate them with speckles in sequence.

CGI has the drawback of a high SR required for the measurement, resulting in a long acquisition time. A CGI system projects many speckle patterns onto an object, then collects the intensities sequentially for ensemble correlation. The requirement for speckle patterns increases immeasurably when the object’s pixel size is large. Many ameliorated techniques have been proposed to minimize the SR, such as orthonormalization method \cite{luo2018orthonormalization,nie2020sub}, Fourier and sequency Walsh-Hadamard speckles~\cite{wang2016fast,zhang2015single,zhang2017hadamard}, and compressive sensing~\cite{katz2009compressive,katkovnik2012compressive}. 

DL-based CGI technique has also shown sub-Nyquist imaging ability. It can retrieve images at a few percentage SRs, which is much lower than other techniques~\cite{lyu2017deep, shimobaba2018computational,barbastathis2019use, wang2019learning,wu2020sub}.  Nevertheless, almost every work uses post-processing techniques, and their adaptive objects are limited to categories from training groups. Therefore, they don’t work or work much worse when objects are outside the training group. In general, these works focus on using DL to suppress the noise fluctuation via matrix restoration or array amelioration algorithm, which does not touch the core concept of ghost imaging. CGI is the linear aggregation of correlation from each pixel where light passes through. To solve the problem fundamentally and universally, we should be direct to the second order correlation.

we conceive that applying DL technique will optimize the cross- and auto-correlation of the speckles. MSE is one of the most frequently appeared evaluators in DL to evaluate picture quality. We therefore choose MSE here as the training loss function for each branch, to compare the CGI results and their ground-truths (other evaluators can also be used as the loss function, as shown in Supplement 1). The MSE is defined as
\begin{equation}
{MSE} = \frac{1}{N_{\mathrm{pixel}}}\sum_{i=1}^{N_{\mathrm{pixel}}}{[\frac{G_\mathrm{i}-X_\mathrm{i}}{\langle G_{(\mathrm{o})}\rangle }]^2}
\label{eq:mse}
\end{equation}
Here, $X$ is the reference matrix calculated by
\begin{equation}
{X_i} =  
\begin{cases}
\langle G_{(\mathrm{o})}\rangle & \text{, Transmission = 1}\\
\langle G_{(\mathrm{b})}\rangle & \text{, Transmission = 0}
\end{cases}
\end{equation}
$G$ represents pixels in the correlation results, $G_{(\mathrm{o})}$ is where the light ought to be transmitted, \textit{i.e.}, the object area, while $G_{(\mathrm{b})}$ is where the light ought to be blocked, \textit{i.e.}, the background area. $\langle\cdots\rangle$ denotes the averaging over all the measurements. $N_{\mathrm{pixel}}$ corresponds to the total pixel number in the speckle patterns ($N_{\mathrm{pixel}}=112\times112$ in our experiments). 

The correlation adjustment method for improving CGI is not limited by training database categories. This one-time training method can let the SR reach an extremely low value. To demonstrate the ability of Speckle-Net, only the MNIST dataset is adopted as training and part of testing images. A total of 60,000 handwritten digits resized to $112\times112$ pixels are used. The optimizer for training process is Stochastic Gradient Descent with Momentum Optimizer (SGDMO) \cite{ruder2016overview}. The momentum of optimizer was set to 0.9 as suggested and weights decay factor was $10^{-3}$ to avoid exploding gradient. After network predicts manipulation on speckles, we utilize training images and patterns to obtain temporary CGIs. The loss function is the MSE between temporary CGIs and original training images, a general loss function for DL problem.  We adopted the mean reduction of each batch as losses since the losses are tremendous for some training images. Then the backwards adjust parameters in the network via manipulation patterns. Generally speaking, the network only relates directly to the speckle patterns instead of training images as in the traditional CNN. As mentioned before, the over-fitting effect is not obvious in our network. Therefore, the network is trained for 200 epochs where the loss stops declining. This program is implemented via Pytorch 1.7.1 and CUDA 11.0 on Python 3.8.5, and we imply GPU-chip NVIDIA GTX1050 for computation acceleration.

The convoluted speckle patterns can then be directly used in the CGI experiment. A typical CGI experiment setup is presented in Fig. ~\ref{fig:setup_CGI}(b). The convoluted speckle patterns from three-step training output are loaded onto a digital micromirror device (DMD). With the illumination from the laser, the speckle patterns are projected to objects, and light passing through the object is collected by the bucket detector (BD). The images can then be retrieved using the standard CGI algorithm.

\section{Characteristics of the Deep-Learned Speckle Patterns}

\begin{figure}[!ht]
    \centering
    \includegraphics[width=\linewidth]{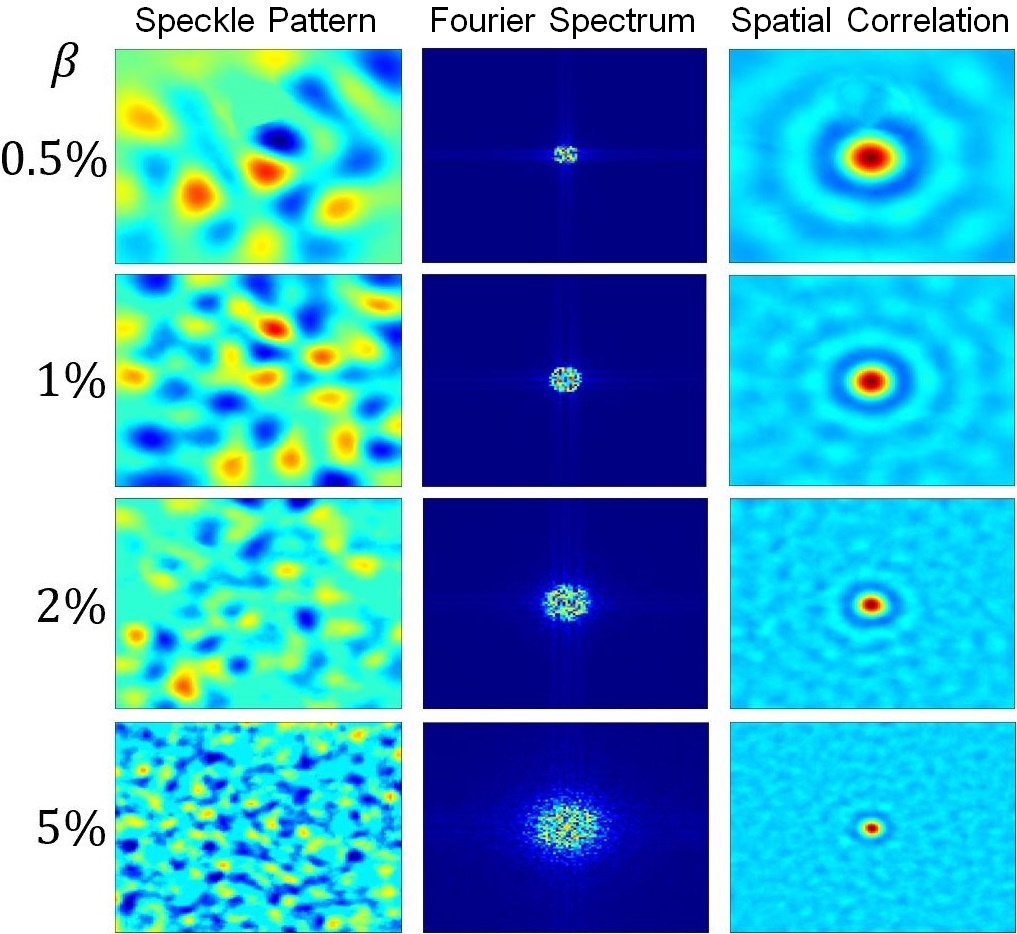}
    \caption{Left column: Typical speckle patterns experiencing three rounds DNN training with SR $\beta=0.5\%, 1\%, 2\%$, and $5\%$; Middle column: The Fourier spectra of corresponding convoluted speckle patterns (The axis values of frequency components $f_x$ and $f_y$ increase from the center to edges.); Right column: The spatial intensity fluctuation correlation distributions of corresponding speckle patterns (The correlation distance increases from the center to edges). }
    \label{fig:compare}
\end{figure}

We choose four different SRs ($\beta=0.5\%$, 1\%, 2\%, and 5\%) for the Speckle-Net training. $\beta$ is defined as $\beta=N_{\mathrm{pattern}}/N_{\mathrm{pixel}}$, where 
$N_{\mathrm{pattern}}$ is the total number of speckle patterns. When $\beta$ is given, the number of kernels $N_{\mathrm k}$ in each layer is settled, $N_{\mathrm k}=\beta N_{\mathrm{pixel}}=N_{\mathrm{pattern}}$. A group of output speckle patterns is given after each round of training with each $\beta$. A typical pink noise speckle pattern~\cite{nie2021noise} is used as the initial pattern. Since the pink noise speckle pattern favors lower spatial frequency components, the training process can converge faster, especially when $\beta$ is small. Three rounds are enough to generate the optimized patterns from the initial pattern via Speckle-Net for all the $\beta$s used in this work, and two rounds are sufficient for smaller $\beta$s (see supplement 1 for detail). In principle, any speckle pattern can be used as the initial input, with possibly extra training (see supplement 1 for detail). 
\begin{figure*}[hbpt!]
    \centering
    \includegraphics[width=\textwidth]{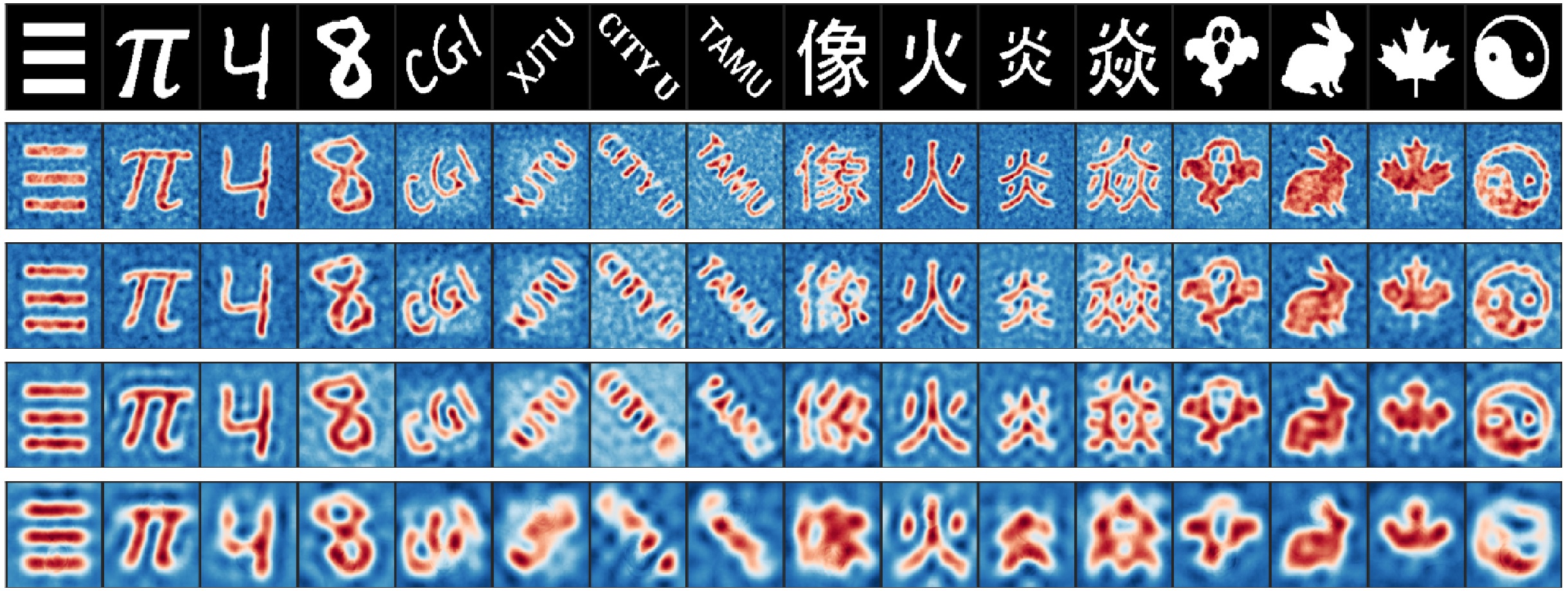}
    \caption{Experimental results of CGI with simple objects (`three lines', $\pi$, digits `4' and `8'), English letters (`CGI', `XJTU', `CITY U', and `TAMU'), Chinese characters (`xi\`{a}ng', `h\v{u}o', `y\'{a}n', and `y\`{a}n', and pictures (`ghost', `rabbit', `leaf', and `Tai Chi') by three rounds deep-learned speckle patterns. From top to bottom: original objects, CGI results with $\beta = 5\%,~2\%,~1\%$, and $0.5\%$, respectively.}
    \label{fig:exp}
\end{figure*}
In Fig.~\ref{fig:compare}, we show the three-round convoluted patterns for various $\beta$ in the first column. The Fourier spectrum distribution and spatial intensity fluctuation correlation distribution $\Gamma^{(2)}(x,y)$ of the patterns are presented in the second and third columns, correspondingly. From Fig.~\ref{fig:compare} we can see that the grain size of the speckle pattern gradually decreases when $\beta$ increases. This is also reflected in the Fourier spectrum distribution, \textit{i.e.}, it concentrates on low spatial frequency when $\beta$ is small, and expands to higher spatial frequencies when $\beta$ increases. Nevertheless, we also notice there are some high frequency components in all the $\beta$ cases, which is also essential for the CGI process.  Now if we check the spatial correlation of the deep-learned speckle patterns, we see that the width of the correlation function is broad when $\beta=0.5\%$, and when $\beta=5\%$, it approaches a delta function. Meanwhile, the background is smoothly distributed, irrespective of $\beta$. This is different than traditional speckle patterns when $\beta$ is small. In the traditional speckle pattern case, the correlation function typically has a significant fluctuation and random distribution in the background due to the lack of ensemble average. Overall, for various $\beta$, the deep-learned speckle patterns always give optimized correlation function which peaked at auto-correlation with certain bandwidth and smoothly distributed cross-correlation background. It is shown in the next section how these unique features will greatly enhance imaging capability at low SRs.

\section{Experimental results}
\subsection{Imaging results with different sampling ratios}

As a demonstration of the effectiveness of deep-learned speckle patterns in CGI system, we conducted a series of measurements using the experimental setup shown in left part of Fig.~\ref{fig:setup_CGI}. The DMD is illuminated by a CW laser, and the deep-learned speckle patterns are sequentially loaded on the DMD then projected to illuminate the object. All objects are $112\times 112$ pixels in size and placed in front of the BD. The BD collects light transmitted from the object. The recorded intensities are then used to make second-order correlations with the corresponding speckle patterns to reconstruct the object. In the experiments, we used our deep-learned speckle patterns from the three rounds training, the with SRs of  0.5\%, 1\%, 2\%, and 5\% (The results from the first round and second round trained speckle patterns are presented in Supplement 1). We adopt four categories of 16 different objects, including simple objects (`three lines', $\pi$, digits `4' and `8'), English letters (`CGI', `XJTU', `CITY U', and `TAMU'), Chinese characters (`xi\`{a}ng', `h\v{u}o', `y\'{a}n', and `y\`{a}n', and pictures (`ghost', `rabbit', `leaf', and `Tai Chi'), for reconstruction. In all the 16 objects, only digits `4' and `8' are from the pattern training dataset. These objects have different sizes, orientations, and complexities to demonstrate the universal adaptability of the deep-learned patterns.  

The main results are shown in Fig.~\ref{fig:exp}. Simple objects such as the simple shape `three lines', Greek letter `$\pi$', digits `4' and `8', and Chinese character `h\v{u}o', can be reconstructed at the SR of only 0.5\%, \textit{i.e.}, only 62 patterns are used for the imaging process. At SR of 1\%, the basic profile can be reconstructed for most of the objects already, and become much more clearer when the SR is 2\%. At the SR of 5\%, all objects can be clearly retrieved. We note here that, when the SR is low, the deep-learned patterns possess higher cross-auto correlation ratio, as shown in Fig.~\ref{fig:compare}. The images generally show higher signal to background ratio but lower resolution. When the SR is high, such as 5\%, the images have much higher resolution. From Fig.~\ref{fig:exp} we can conclude that all the objects with different complexity can be reconstructed with high visibility and low noise fluctuation in the background. This boost the deep-learned speckle patterns' applicability in extremely low sampling ranges, which might be useful in moving object capture and dynamic imaging systems.

\subsection{Imaging results under different noise conditions}
\begin{figure*}[!hpt]
    \centering
    \includegraphics[width=\linewidth]{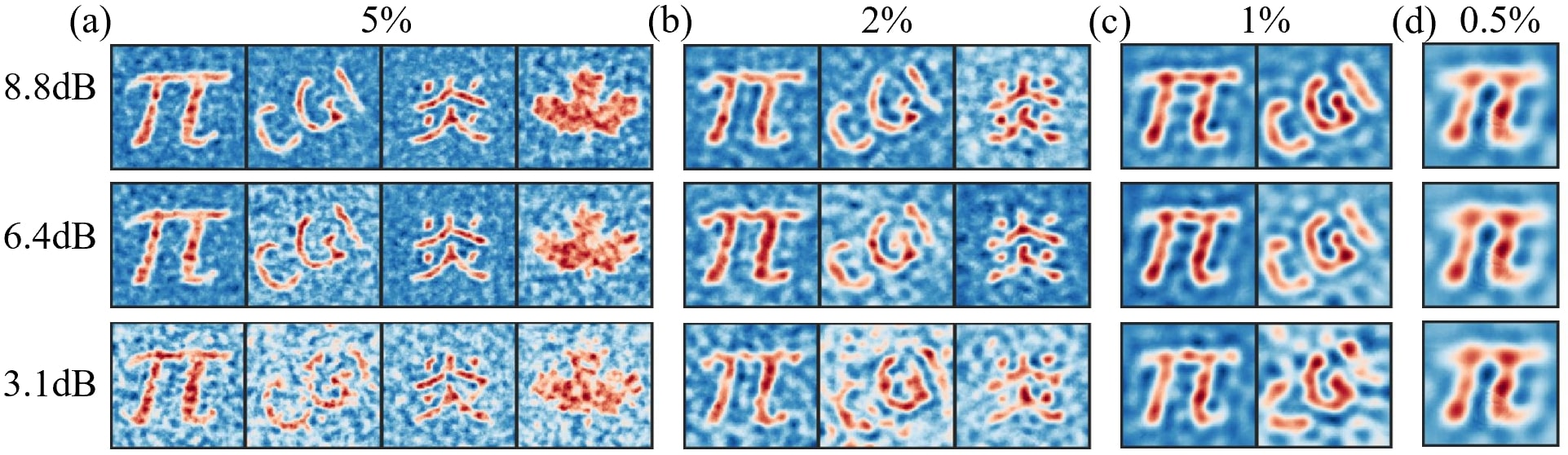}
    \caption{Experimental results of CGI using Deep-learned speckles with different noise levels labelled in the left column. (a) CGI results with $\beta=5\%$, (b) CGI results with $\beta=2\%$, (c) CGI results with $\beta=1\%$, and (d) CGI results with $\beta=0.5\%$. }
    \label{fig:exp_noise}
\end{figure*}

Another advantage of the deep-learned speckle pattern is that, the optimized auto- and cross-correlation enables its noise-robust feature meanwhile possesses sufficient spatial resolution. To demonstrate the ability of imaging under noisy interference of the deep-learned patterns, we perform a series of measurements of four objects under different noise levels. We choose the four objects from our four catalogs: Greek letter `$\pi$', English letters `CGI', Chinese character `y\'{a}n', and picture `leaf'. The noise level is represented by the Signal-to-noise ratio (SNR) value. The SNR in logarithmic decibel scale is defined as
\begin{equation}
\mathrm{SNR} =10 \log\frac{P_{\mathrm{s}}}{P_{\mathrm{b}}},
\end{equation}
where  $P_{\mathrm{s}}$ is the average intensity in each signal pixel and $P_{\mathrm{b}}$ is the average intensity in the noise background. Here we choose three different SNRs: $8.8 \mathrm{dB}$, $6.4 \mathrm{dB}$, and $3.1 \mathrm{dB}$.

The results are shown in Fig.~\ref{fig:exp_noise}. It is clearly seen that at $8.8 \mathrm{dB}$, all the images can be retrieved at all different SRs. When the SNR is $6.4 \mathrm{dB}$, some images start to show a noisy background. Nevertheless, all the objects can still be clearly identified. When the SNR is $3.1 \mathrm{dB}$, which can be considered very noisy, most of the objects can still be identified. We also notice that, speckle patterns with lower SR are more robust to noise interference. Take the Greek letter '$\pi$' for example, although it can be clearly imaged at $3.1 \mathrm{dB}$ when the SR is 5\%, noticeable background noise exists in the resulting image. At 2\% SR, the background noise starts to degrade.  When the SR is at 1\% or 0.5\%, the background is almost smooth, and we see nearly no difference between the results at the three noise levels. 

The noise-robust feature results from the optimized cross-auto correlation ratio for each SR. At the extremely low SR 1\% and 0.5\%, the cross-correlation is emphasized to enhance the SNR and suppress the fluctuations due to limited sampling. Therefore, the deep-learned speckle patterns are feasible to apply in noisy environments.

\section{Conclusion and Discussion}

In summary, we propose a speckle pattern generation scheme, Speckle-Net, through the use of Deep Learning algorithms and concepts to achieve the desired feature. Then, we select the standard CGI algorithm as our objective for the loss function and use this method to generate speckle patterns for CGI. We demonstrate experimentally that the deep-learned speckle pattern can be used for the standard CGI measurement, enhance the imaging efficiency, and be robust to noise. The method is unique and superior to the traditional CGI and deep-learning-based CGI focusing on image amelioration or imaging algorithms. The multi-branched Speckle-Net gives flexibility in finding globally optimal solutions while reducing training time. Also, since the learning process only focuses on the speckle patterns, it can be used for other speckle illumination systems by changing the objective in the loss function. Even though the network is trained only using the MNIST digit dataset, the resulting speckle patterns can retrieve images for simple objects with an extremely low SR (0.5\%) and complicated objects with only a 5\% SR. Additionally, a CGI system with deep-learned speckle patterns is insensitive to noise interference.

Although a particular example, \textit{i.e.}, the CGI is demonstrated in this work, in the long term, we believe the pioneering work boosts a closer connection between DL and speckle pattern generation, which will pave the way for broader and practical exploitation of ghost imaging and other applications. For instance, begin from light field propagation function, we can let the neural network to learn and create an unique phase distributions according to the contrast parameter while propagation, which may manipulate a kind of non-diffractive speckle pattern in photonics lattice applications~\cite{liu2021generation}. In addition, other structures such as U-net~\cite{ronneberger2015u}, recurrent neural network (RNN)~\cite{connor1994recurrent}, transformer~\cite{jaderberg2015spatial,vaswani2017attention}, {\it etc.}, can be similarly explored and modified to generate aimed speckle patterns. For example, the time-dependent RNN and transformer can be modified similarly as what we do on CNN to make other types of Speckle-Net which can fabricate time-dependent speckle patterns according to the instant feedback and demand of systems during the measurement. Specifically, the $n$-th illumination pattern can be generated from patterns and results with $n-1$ sampling number.

\section*{Funding.}
Air Force Office of Scientific Research (Award No. FA9550-20-1-0366 DEF), Office of Naval Research (Award No. N00014-20-1-2184), Robert A. Welch Foundation (Grant No. A-1261), and National Science Foundation (Grant No. PHY-2013771).

\section*{Data Availability.} The experiment data and convoluted speckle patterns in this article are available upon reasonable request from the authors. The Speckle-Net and initial patterns can be found at \url{https://github.com/XJTU-TAMU-CGI/PatternDL}.

\section*{Disclosures.} The authors declare no conflicts of interest.


\bibliography{DLpattern}

\begin{thebibliography}{55}%
\makeatletter
\providecommand \@ifxundefined [1]{%
 \@ifx{#1\undefined}
}%
\providecommand \@ifnum [1]{%
 \ifnum #1\expandafter \@firstoftwo
 \else \expandafter \@secondoftwo
 \fi
}%
\providecommand \@ifx [1]{%
 \ifx #1\expandafter \@firstoftwo
 \else \expandafter \@secondoftwo
 \fi
}%
\providecommand \natexlab [1]{#1}%
\providecommand \enquote  [1]{``#1''}%
\providecommand \bibnamefont  [1]{#1}%
\providecommand \bibfnamefont [1]{#1}%
\providecommand \citenamefont [1]{#1}%
\providecommand \href@noop [0]{\@secondoftwo}%
\providecommand \href [0]{\begingroup \@sanitize@url \@href}%
\providecommand \@href[1]{\@@startlink{#1}\@@href}%
\providecommand \@@href[1]{\endgroup#1\@@endlink}%
\providecommand \@sanitize@url [0]{\catcode `\\12\catcode `\$12\catcode
  `\&12\catcode `\#12\catcode `\^12\catcode `\_12\catcode `\%12\relax}%
\providecommand \@@startlink[1]{}%
\providecommand \@@endlink[0]{}%
\providecommand \url  [0]{\begingroup\@sanitize@url \@url }%
\providecommand \@url [1]{\endgroup\@href {#1}{\urlprefix }}%
\providecommand \urlprefix  [0]{URL }%
\providecommand \Eprint [0]{\href }%
\providecommand \doibase [0]{http://dx.doi.org/}%
\providecommand \selectlanguage [0]{\@gobble}%
\providecommand \bibinfo  [0]{\@secondoftwo}%
\providecommand \bibfield  [0]{\@secondoftwo}%
\providecommand \translation [1]{[#1]}%
\providecommand \BibitemOpen [0]{}%
\providecommand \bibitemStop [0]{}%
\providecommand \bibitemNoStop [0]{.\EOS\space}%
\providecommand \EOS [0]{\spacefactor3000\relax}%
\providecommand \BibitemShut  [1]{\csname bibitem#1\endcsname}%
\let\auto@bib@innerbib\@empty
\bibitem [{\citenamefont {Pine}\ \emph {et~al.}(1988)\citenamefont {Pine},
  \citenamefont {Weitz}, \citenamefont {Chaikin},\ and\ \citenamefont
  {Herbolzheimer}}]{pine1988diffusing}%
  \BibitemOpen
  \bibfield  {author} {\bibinfo {author} {\bibfnamefont {David~J}\ \bibnamefont
  {Pine}}, \bibinfo {author} {\bibfnamefont {David~A}\ \bibnamefont {Weitz}},
  \bibinfo {author} {\bibfnamefont {Paul~M}\ \bibnamefont {Chaikin}}, \ and\
  \bibinfo {author} {\bibfnamefont {Eric}\ \bibnamefont {Herbolzheimer}},\
  }\bibfield  {title} {\enquote {\bibinfo {title} {Diffusing wave
  spectroscopy},}\ }\href@noop {} {\bibfield  {journal} {\bibinfo  {journal}
  {Physical Review Letters}\ }\textbf {\bibinfo {volume} {60}},\ \bibinfo
  {pages} {1134} (\bibinfo {year} {1988})}\BibitemShut {NoStop}%
\bibitem [{\citenamefont {Li}\ \emph {et~al.}(2020)\citenamefont {Li},
  \citenamefont {Li}, \citenamefont {Peng},\ and\ \citenamefont
  {Agarwal}}]{li2020photon}%
  \BibitemOpen
  \bibfield  {author} {\bibinfo {author} {\bibfnamefont {Sheng-Wen}\
  \bibnamefont {Li}}, \bibinfo {author} {\bibfnamefont {Fu}~\bibnamefont {Li}},
  \bibinfo {author} {\bibfnamefont {Tao}\ \bibnamefont {Peng}}, \ and\ \bibinfo
  {author} {\bibfnamefont {GS}~\bibnamefont {Agarwal}},\ }\bibfield  {title}
  {\enquote {\bibinfo {title} {Photon statistics of quantum light on scattering
  from rotating ground glass},}\ }\href@noop {} {\bibfield  {journal} {\bibinfo
   {journal} {Physical Review A}\ }\textbf {\bibinfo {volume} {101}},\ \bibinfo
  {pages} {063806} (\bibinfo {year} {2020})}\BibitemShut {NoStop}%
\bibitem [{\citenamefont {Goodman}(1975)}]{goodman1975statistical}%
  \BibitemOpen
  \bibfield  {author} {\bibinfo {author} {\bibfnamefont {Joseph~W}\
  \bibnamefont {Goodman}},\ }\bibfield  {title} {\enquote {\bibinfo {title}
  {Statistical properties of laser speckle patterns},}\ }in\ \href@noop {}
  {\emph {\bibinfo {booktitle} {Laser Speckle and Related Phenomena}}}\
  (\bibinfo  {publisher} {Springer},\ \bibinfo {year} {1975})\ pp.\ \bibinfo
  {pages} {9--75}\BibitemShut {NoStop}%
\bibitem [{\citenamefont {Zanette}\ \emph {et~al.}(2014)\citenamefont
  {Zanette}, \citenamefont {Zhou}, \citenamefont {Burvall}, \citenamefont
  {Lundstr{\"o}m}, \citenamefont {Larsson}, \citenamefont {Zdora},
  \citenamefont {Thibault}, \citenamefont {Pfeiffer},\ and\ \citenamefont
  {Hertz}}]{zanette2014speckle}%
  \BibitemOpen
  \bibfield  {author} {\bibinfo {author} {\bibfnamefont {I}~\bibnamefont
  {Zanette}}, \bibinfo {author} {\bibfnamefont {Tunhe}\ \bibnamefont {Zhou}},
  \bibinfo {author} {\bibfnamefont {Anna}\ \bibnamefont {Burvall}}, \bibinfo
  {author} {\bibfnamefont {Ulf}\ \bibnamefont {Lundstr{\"o}m}}, \bibinfo
  {author} {\bibfnamefont {Daniel~H}\ \bibnamefont {Larsson}}, \bibinfo
  {author} {\bibfnamefont {M}~\bibnamefont {Zdora}}, \bibinfo {author}
  {\bibfnamefont {P}~\bibnamefont {Thibault}}, \bibinfo {author} {\bibfnamefont
  {Franz}\ \bibnamefont {Pfeiffer}}, \ and\ \bibinfo {author} {\bibfnamefont
  {Hans~M}\ \bibnamefont {Hertz}},\ }\bibfield  {title} {\enquote {\bibinfo
  {title} {Speckle-based x-ray phase-contrast and dark-field imaging with a
  laboratory source},}\ }\href@noop {} {\bibfield  {journal} {\bibinfo
  {journal} {Physical Review Letters}\ }\textbf {\bibinfo {volume} {112}},\
  \bibinfo {pages} {253903} (\bibinfo {year} {2014})}\BibitemShut {NoStop}%
\bibitem [{\citenamefont {Wang}\ and\ \citenamefont
  {Genack}(2011)}]{wang2011transport}%
  \BibitemOpen
  \bibfield  {author} {\bibinfo {author} {\bibfnamefont {Jing}\ \bibnamefont
  {Wang}}\ and\ \bibinfo {author} {\bibfnamefont {Azriel~Z}\ \bibnamefont
  {Genack}},\ }\bibfield  {title} {\enquote {\bibinfo {title} {Transport
  through modes in random media},}\ }\href@noop {} {\bibfield  {journal}
  {\bibinfo  {journal} {Nature}\ }\textbf {\bibinfo {volume} {471}},\ \bibinfo
  {pages} {345--348} (\bibinfo {year} {2011})}\BibitemShut {NoStop}%
\bibitem [{\citenamefont {Olivieri}\ \emph {et~al.}(2020)\citenamefont
  {Olivieri}, \citenamefont {Gongora}, \citenamefont {Peters}, \citenamefont
  {Cecconi}, \citenamefont {Cutrona}, \citenamefont {Tunesi}, \citenamefont
  {Tucker}, \citenamefont {Pasquazi},\ and\ \citenamefont
  {Peccianti}}]{olivieri2020hyperspectral}%
  \BibitemOpen
  \bibfield  {author} {\bibinfo {author} {\bibfnamefont {Luana}\ \bibnamefont
  {Olivieri}}, \bibinfo {author} {\bibfnamefont {Juan S~Totero}\ \bibnamefont
  {Gongora}}, \bibinfo {author} {\bibfnamefont {Luke}\ \bibnamefont {Peters}},
  \bibinfo {author} {\bibfnamefont {Vittorio}\ \bibnamefont {Cecconi}},
  \bibinfo {author} {\bibfnamefont {Antonio}\ \bibnamefont {Cutrona}}, \bibinfo
  {author} {\bibfnamefont {Jacob}\ \bibnamefont {Tunesi}}, \bibinfo {author}
  {\bibfnamefont {Robyn}\ \bibnamefont {Tucker}}, \bibinfo {author}
  {\bibfnamefont {Alessia}\ \bibnamefont {Pasquazi}}, \ and\ \bibinfo {author}
  {\bibfnamefont {Marco}\ \bibnamefont {Peccianti}},\ }\bibfield  {title}
  {\enquote {\bibinfo {title} {Hyperspectral terahertz microscopy via nonlinear
  ghost imaging},}\ }\href@noop {} {\bibfield  {journal} {\bibinfo  {journal}
  {Optica}\ }\textbf {\bibinfo {volume} {7}},\ \bibinfo {pages} {186--191}
  (\bibinfo {year} {2020})}\BibitemShut {NoStop}%
\bibitem [{\citenamefont {Valley}\ \emph {et~al.}(2016)\citenamefont {Valley},
  \citenamefont {Sefler},\ and\ \citenamefont {Shaw}}]{valley2016multimode}%
  \BibitemOpen
  \bibfield  {author} {\bibinfo {author} {\bibfnamefont {George~C}\
  \bibnamefont {Valley}}, \bibinfo {author} {\bibfnamefont {George~A}\
  \bibnamefont {Sefler}}, \ and\ \bibinfo {author} {\bibfnamefont {T~Justin}\
  \bibnamefont {Shaw}},\ }\bibfield  {title} {\enquote {\bibinfo {title}
  {Multimode waveguide speckle patterns for compressive sensing},}\ }\href@noop
  {} {\bibfield  {journal} {\bibinfo  {journal} {Optics Letters}\ }\textbf
  {\bibinfo {volume} {41}},\ \bibinfo {pages} {2529--2532} (\bibinfo {year}
  {2016})}\BibitemShut {NoStop}%
\bibitem [{\citenamefont {Redding}\ \emph
  {et~al.}(2013{\natexlab{a}})\citenamefont {Redding}, \citenamefont {Popoff},\
  and\ \citenamefont {Cao}}]{redding2013all}%
  \BibitemOpen
  \bibfield  {author} {\bibinfo {author} {\bibfnamefont {Brandon}\ \bibnamefont
  {Redding}}, \bibinfo {author} {\bibfnamefont {Sebastien~M}\ \bibnamefont
  {Popoff}}, \ and\ \bibinfo {author} {\bibfnamefont {Hui}\ \bibnamefont
  {Cao}},\ }\bibfield  {title} {\enquote {\bibinfo {title} {All-fiber
  spectrometer based on speckle pattern reconstruction},}\ }\href@noop {}
  {\bibfield  {journal} {\bibinfo  {journal} {Optics Express}\ }\textbf
  {\bibinfo {volume} {21}},\ \bibinfo {pages} {6584--6600} (\bibinfo {year}
  {2013}{\natexlab{a}})}\BibitemShut {NoStop}%
\bibitem [{\citenamefont {Strudley}\ \emph {et~al.}(2013)\citenamefont
  {Strudley}, \citenamefont {Zehender}, \citenamefont {Blejean}, \citenamefont
  {Bakkers},\ and\ \citenamefont {Muskens}}]{strudley2013mesoscopic}%
  \BibitemOpen
  \bibfield  {author} {\bibinfo {author} {\bibfnamefont {Tom}\ \bibnamefont
  {Strudley}}, \bibinfo {author} {\bibfnamefont {Tilman}\ \bibnamefont
  {Zehender}}, \bibinfo {author} {\bibfnamefont {Claire}\ \bibnamefont
  {Blejean}}, \bibinfo {author} {\bibfnamefont {Erik~PAM}\ \bibnamefont
  {Bakkers}}, \ and\ \bibinfo {author} {\bibfnamefont {Otto~L}\ \bibnamefont
  {Muskens}},\ }\bibfield  {title} {\enquote {\bibinfo {title} {Mesoscopic
  light transport by very strong collective multiple scattering in nanowire
  mats},}\ }\href@noop {} {\bibfield  {journal} {\bibinfo  {journal} {Nature
  Photonics}\ }\textbf {\bibinfo {volume} {7}},\ \bibinfo {pages} {413--418}
  (\bibinfo {year} {2013})}\BibitemShut {NoStop}%
\bibitem [{\citenamefont {Redding}\ \emph
  {et~al.}(2013{\natexlab{b}})\citenamefont {Redding}, \citenamefont {Liew},
  \citenamefont {Sarma},\ and\ \citenamefont {Cao}}]{redding2013compact}%
  \BibitemOpen
  \bibfield  {author} {\bibinfo {author} {\bibfnamefont {Brandon}\ \bibnamefont
  {Redding}}, \bibinfo {author} {\bibfnamefont {Seng~Fatt}\ \bibnamefont
  {Liew}}, \bibinfo {author} {\bibfnamefont {Raktim}\ \bibnamefont {Sarma}}, \
  and\ \bibinfo {author} {\bibfnamefont {Hui}\ \bibnamefont {Cao}},\ }\bibfield
   {title} {\enquote {\bibinfo {title} {Compact spectrometer based on a
  disordered photonic chip},}\ }\href@noop {} {\bibfield  {journal} {\bibinfo
  {journal} {Nature Photonics}\ }\textbf {\bibinfo {volume} {7}},\ \bibinfo
  {pages} {746--751} (\bibinfo {year} {2013}{\natexlab{b}})}\BibitemShut
  {NoStop}%
\bibitem [{\citenamefont {Ventalon}\ and\ \citenamefont
  {Mertz}(2006)}]{ventalon2006dynamic}%
  \BibitemOpen
  \bibfield  {author} {\bibinfo {author} {\bibfnamefont {Cathie}\ \bibnamefont
  {Ventalon}}\ and\ \bibinfo {author} {\bibfnamefont {Jerome}\ \bibnamefont
  {Mertz}},\ }\bibfield  {title} {\enquote {\bibinfo {title} {Dynamic speckle
  illumination microscopy with translated versus randomized speckle
  patterns},}\ }\href@noop {} {\bibfield  {journal} {\bibinfo  {journal}
  {Optics Express}\ }\textbf {\bibinfo {volume} {14}},\ \bibinfo {pages}
  {7198--7209} (\bibinfo {year} {2006})}\BibitemShut {NoStop}%
\bibitem [{\citenamefont {Mertz}(2011)}]{mertz2011optical}%
  \BibitemOpen
  \bibfield  {author} {\bibinfo {author} {\bibfnamefont {Jerome}\ \bibnamefont
  {Mertz}},\ }\bibfield  {title} {\enquote {\bibinfo {title} {Optical
  sectioning microscopy with planar or structured illumination},}\ }\href@noop
  {} {\bibfield  {journal} {\bibinfo  {journal} {Nature methods}\ }\textbf
  {\bibinfo {volume} {8}},\ \bibinfo {pages} {811} (\bibinfo {year}
  {2011})}\BibitemShut {NoStop}%
\bibitem [{\citenamefont {Mudry}\ \emph {et~al.}(2012)\citenamefont {Mudry},
  \citenamefont {Belkebir}, \citenamefont {Girard}, \citenamefont {Savatier},
  \citenamefont {Le~Moal}, \citenamefont {Nicoletti}, \citenamefont {Allain},\
  and\ \citenamefont {Sentenac}}]{mudry2012structured}%
  \BibitemOpen
  \bibfield  {author} {\bibinfo {author} {\bibfnamefont {Emeric}\ \bibnamefont
  {Mudry}}, \bibinfo {author} {\bibfnamefont {Kamal}\ \bibnamefont {Belkebir}},
  \bibinfo {author} {\bibfnamefont {J}~\bibnamefont {Girard}}, \bibinfo
  {author} {\bibfnamefont {Julien}\ \bibnamefont {Savatier}}, \bibinfo {author}
  {\bibfnamefont {Emmeran}\ \bibnamefont {Le~Moal}}, \bibinfo {author}
  {\bibfnamefont {C}~\bibnamefont {Nicoletti}}, \bibinfo {author}
  {\bibfnamefont {Marc}\ \bibnamefont {Allain}}, \ and\ \bibinfo {author}
  {\bibfnamefont {Anne}\ \bibnamefont {Sentenac}},\ }\bibfield  {title}
  {\enquote {\bibinfo {title} {Structured illumination microscopy using unknown
  speckle patterns},}\ }\href@noop {} {\bibfield  {journal} {\bibinfo
  {journal} {Nature Photonics}\ }\textbf {\bibinfo {volume} {6}},\ \bibinfo
  {pages} {312--315} (\bibinfo {year} {2012})}\BibitemShut {NoStop}%
\bibitem [{\citenamefont {Nakadate}\ and\ \citenamefont
  {Saito}(1985)}]{nakadate1985fringe}%
  \BibitemOpen
  \bibfield  {author} {\bibinfo {author} {\bibfnamefont {Suezou}\ \bibnamefont
  {Nakadate}}\ and\ \bibinfo {author} {\bibfnamefont {Hiroyoshi}\ \bibnamefont
  {Saito}},\ }\bibfield  {title} {\enquote {\bibinfo {title} {Fringe scanning
  speckle-pattern interferometry},}\ }\href@noop {} {\bibfield  {journal}
  {\bibinfo  {journal} {Applied Optics}\ }\textbf {\bibinfo {volume} {24}},\
  \bibinfo {pages} {2172--2180} (\bibinfo {year} {1985})}\BibitemShut {NoStop}%
\bibitem [{\citenamefont {Yilmaz}\ \emph {et~al.}(2015)\citenamefont {Yilmaz},
  \citenamefont {van Putten}, \citenamefont {Bertolotti}, \citenamefont
  {Lagendijk}, \citenamefont {Vos},\ and\ \citenamefont
  {Mosk}}]{yilmaz2015speckle}%
  \BibitemOpen
  \bibfield  {author} {\bibinfo {author} {\bibfnamefont {Hasan}\ \bibnamefont
  {Yilmaz}}, \bibinfo {author} {\bibfnamefont {Elbert~G}\ \bibnamefont {van
  Putten}}, \bibinfo {author} {\bibfnamefont {Jacopo}\ \bibnamefont
  {Bertolotti}}, \bibinfo {author} {\bibfnamefont {Ad}~\bibnamefont
  {Lagendijk}}, \bibinfo {author} {\bibfnamefont {Willem~L}\ \bibnamefont
  {Vos}}, \ and\ \bibinfo {author} {\bibfnamefont {Allard~P}\ \bibnamefont
  {Mosk}},\ }\bibfield  {title} {\enquote {\bibinfo {title} {Speckle
  correlation resolution enhancement of wide-field fluorescence imaging},}\
  }\href@noop {} {\bibfield  {journal} {\bibinfo  {journal} {Optica}\ }\textbf
  {\bibinfo {volume} {2}},\ \bibinfo {pages} {424--429} (\bibinfo {year}
  {2015})}\BibitemShut {NoStop}%
\bibitem [{\citenamefont {Pascucci}\ \emph {et~al.}(2016)\citenamefont
  {Pascucci}, \citenamefont {Tessier}, \citenamefont {Emiliani},\ and\
  \citenamefont {Guillon}}]{pascucci2016superresolution}%
  \BibitemOpen
  \bibfield  {author} {\bibinfo {author} {\bibfnamefont {Marco}\ \bibnamefont
  {Pascucci}}, \bibinfo {author} {\bibfnamefont {Gilles}\ \bibnamefont
  {Tessier}}, \bibinfo {author} {\bibfnamefont {Valentina}\ \bibnamefont
  {Emiliani}}, \ and\ \bibinfo {author} {\bibfnamefont {Marc}\ \bibnamefont
  {Guillon}},\ }\bibfield  {title} {\enquote {\bibinfo {title} {Superresolution
  imaging of optical vortices in a speckle pattern},}\ }\href@noop {}
  {\bibfield  {journal} {\bibinfo  {journal} {Physical Review Letters}\
  }\textbf {\bibinfo {volume} {116}},\ \bibinfo {pages} {093904} (\bibinfo
  {year} {2016})}\BibitemShut {NoStop}%
\bibitem [{\citenamefont {McGehee}\ \emph {et~al.}(2013)\citenamefont
  {McGehee}, \citenamefont {Kondov}, \citenamefont {Xu}, \citenamefont
  {Zirbel},\ and\ \citenamefont {DeMarco}}]{mcgehee2013three}%
  \BibitemOpen
  \bibfield  {author} {\bibinfo {author} {\bibfnamefont {WR}~\bibnamefont
  {McGehee}}, \bibinfo {author} {\bibfnamefont {SS}~\bibnamefont {Kondov}},
  \bibinfo {author} {\bibfnamefont {W}~\bibnamefont {Xu}}, \bibinfo {author}
  {\bibfnamefont {JJ}~\bibnamefont {Zirbel}}, \ and\ \bibinfo {author}
  {\bibfnamefont {B}~\bibnamefont {DeMarco}},\ }\bibfield  {title} {\enquote
  {\bibinfo {title} {Three-dimensional anderson localization in variable scale
  disorder},}\ }\href@noop {} {\bibfield  {journal} {\bibinfo  {journal}
  {Physical review letters}\ }\textbf {\bibinfo {volume} {111}},\ \bibinfo
  {pages} {145303} (\bibinfo {year} {2013})}\BibitemShut {NoStop}%
\bibitem [{\citenamefont {Delande}\ and\ \citenamefont
  {Orso}(2014)}]{delande2014mobility}%
  \BibitemOpen
  \bibfield  {author} {\bibinfo {author} {\bibfnamefont {Dominique}\
  \bibnamefont {Delande}}\ and\ \bibinfo {author} {\bibfnamefont {Giuliano}\
  \bibnamefont {Orso}},\ }\bibfield  {title} {\enquote {\bibinfo {title}
  {Mobility edge for cold atoms in laser speckle potentials},}\ }\href@noop {}
  {\bibfield  {journal} {\bibinfo  {journal} {Physical review letters}\
  }\textbf {\bibinfo {volume} {113}},\ \bibinfo {pages} {060601} (\bibinfo
  {year} {2014})}\BibitemShut {NoStop}%
\bibitem [{\citenamefont {Fratini}\ and\ \citenamefont
  {Pilati}(2015)}]{fratini2015anderson}%
  \BibitemOpen
  \bibfield  {author} {\bibinfo {author} {\bibfnamefont {Elisa}\ \bibnamefont
  {Fratini}}\ and\ \bibinfo {author} {\bibfnamefont {Sebastiano}\ \bibnamefont
  {Pilati}},\ }\bibfield  {title} {\enquote {\bibinfo {title} {Anderson
  localization of matter waves in quantum-chaos theory},}\ }\href@noop {}
  {\bibfield  {journal} {\bibinfo  {journal} {Physical Review A}\ }\textbf
  {\bibinfo {volume} {91}},\ \bibinfo {pages} {061601} (\bibinfo {year}
  {2015})}\BibitemShut {NoStop}%
\bibitem [{\citenamefont {Liu}\ \emph {et~al.}(2021)\citenamefont {Liu},
  \citenamefont {Qing}, \citenamefont {Zhao}, \citenamefont {Zhang},
  \citenamefont {Gao}, \citenamefont {Chen},\ and\ \citenamefont
  {Li}}]{liu2021generation}%
  \BibitemOpen
  \bibfield  {author} {\bibinfo {author} {\bibfnamefont {Ruifeng}\ \bibnamefont
  {Liu}}, \bibinfo {author} {\bibfnamefont {Bingcheng}\ \bibnamefont {Qing}},
  \bibinfo {author} {\bibfnamefont {Shupeng}\ \bibnamefont {Zhao}}, \bibinfo
  {author} {\bibfnamefont {Pei}\ \bibnamefont {Zhang}}, \bibinfo {author}
  {\bibfnamefont {Hong}\ \bibnamefont {Gao}}, \bibinfo {author} {\bibfnamefont
  {Shouqian}\ \bibnamefont {Chen}}, \ and\ \bibinfo {author} {\bibfnamefont
  {Fuli}\ \bibnamefont {Li}},\ }\bibfield  {title} {\enquote {\bibinfo {title}
  {Generation of non-rayleigh nondiffracting speckles},}\ }\href@noop {}
  {\bibfield  {journal} {\bibinfo  {journal} {Physical Review Letters}\
  }\textbf {\bibinfo {volume} {127}},\ \bibinfo {pages} {180601} (\bibinfo
  {year} {2021})}\BibitemShut {NoStop}%
\bibitem [{\citenamefont {Bennink}\ \emph {et~al.}(2002)\citenamefont
  {Bennink}, \citenamefont {Bentley},\ and\ \citenamefont
  {Boyd}}]{bennink2002two}%
  \BibitemOpen
  \bibfield  {author} {\bibinfo {author} {\bibfnamefont {Ryan~S}\ \bibnamefont
  {Bennink}}, \bibinfo {author} {\bibfnamefont {Sean~J}\ \bibnamefont
  {Bentley}}, \ and\ \bibinfo {author} {\bibfnamefont {Robert~W}\ \bibnamefont
  {Boyd}},\ }\bibfield  {title} {\enquote {\bibinfo {title} {“two-photon”
  coincidence imaging with a classical source},}\ }\href@noop {} {\bibfield
  {journal} {\bibinfo  {journal} {Physical Review Letters}\ }\textbf {\bibinfo
  {volume} {89}},\ \bibinfo {pages} {113601} (\bibinfo {year}
  {2002})}\BibitemShut {NoStop}%
\bibitem [{\citenamefont {Chen}\ \emph {et~al.}(2009)\citenamefont {Chen},
  \citenamefont {Liu}, \citenamefont {Luo},\ and\ \citenamefont
  {Wu}}]{chen2009lensless}%
  \BibitemOpen
  \bibfield  {author} {\bibinfo {author} {\bibfnamefont {Xi-Hao}\ \bibnamefont
  {Chen}}, \bibinfo {author} {\bibfnamefont {Qian}\ \bibnamefont {Liu}},
  \bibinfo {author} {\bibfnamefont {Kai-Hong}\ \bibnamefont {Luo}}, \ and\
  \bibinfo {author} {\bibfnamefont {Ling-An}\ \bibnamefont {Wu}},\ }\bibfield
  {title} {\enquote {\bibinfo {title} {Lensless ghost imaging with true thermal
  light},}\ }\href@noop {} {\bibfield  {journal} {\bibinfo  {journal} {Optics
  Letters}\ }\textbf {\bibinfo {volume} {34}},\ \bibinfo {pages} {695--697}
  (\bibinfo {year} {2009})}\BibitemShut {NoStop}%
\bibitem [{\citenamefont {Valencia}\ \emph {et~al.}(2005)\citenamefont
  {Valencia}, \citenamefont {Scarcelli}, \citenamefont {D’Angelo},\ and\
  \citenamefont {Shih}}]{valencia2005two}%
  \BibitemOpen
  \bibfield  {author} {\bibinfo {author} {\bibfnamefont {Alejandra}\
  \bibnamefont {Valencia}}, \bibinfo {author} {\bibfnamefont {Giuliano}\
  \bibnamefont {Scarcelli}}, \bibinfo {author} {\bibfnamefont {Milena}\
  \bibnamefont {D’Angelo}}, \ and\ \bibinfo {author} {\bibfnamefont {Yanhua}\
  \bibnamefont {Shih}},\ }\bibfield  {title} {\enquote {\bibinfo {title}
  {Two-photon imaging with thermal light},}\ }\href@noop {} {\bibfield
  {journal} {\bibinfo  {journal} {Physical Review Letters}\ }\textbf {\bibinfo
  {volume} {94}},\ \bibinfo {pages} {063601} (\bibinfo {year}
  {2005})}\BibitemShut {NoStop}%
\bibitem [{\citenamefont {Shapiro}(2008)}]{shapiro2008computational}%
  \BibitemOpen
  \bibfield  {author} {\bibinfo {author} {\bibfnamefont {Jeffrey~H}\
  \bibnamefont {Shapiro}},\ }\bibfield  {title} {\enquote {\bibinfo {title}
  {Computational ghost imaging},}\ }\href@noop {} {\bibfield  {journal}
  {\bibinfo  {journal} {Physical Review A}\ }\textbf {\bibinfo {volume} {78}},\
  \bibinfo {pages} {061802} (\bibinfo {year} {2008})}\BibitemShut {NoStop}%
\bibitem [{\citenamefont {Bromberg}\ and\ \citenamefont
  {Cao}(2014)}]{bromberg2014generating}%
  \BibitemOpen
  \bibfield  {author} {\bibinfo {author} {\bibfnamefont {Yaron}\ \bibnamefont
  {Bromberg}}\ and\ \bibinfo {author} {\bibfnamefont {Hui}\ \bibnamefont
  {Cao}},\ }\bibfield  {title} {\enquote {\bibinfo {title} {Generating
  non-rayleigh speckles with tailored intensity statistics},}\ }\href@noop {}
  {\bibfield  {journal} {\bibinfo  {journal} {Physical Review Letters}\
  }\textbf {\bibinfo {volume} {112}},\ \bibinfo {pages} {213904} (\bibinfo
  {year} {2014})}\BibitemShut {NoStop}%
\bibitem [{\citenamefont {Kondakci}\ \emph {et~al.}(2016)\citenamefont
  {Kondakci}, \citenamefont {Szameit}, \citenamefont {Abouraddy}, \citenamefont
  {Christodoulides},\ and\ \citenamefont {Saleh}}]{kondakci2016sub}%
  \BibitemOpen
  \bibfield  {author} {\bibinfo {author} {\bibfnamefont {H~Esat}\ \bibnamefont
  {Kondakci}}, \bibinfo {author} {\bibfnamefont {Alexander}\ \bibnamefont
  {Szameit}}, \bibinfo {author} {\bibfnamefont {Ayman~F}\ \bibnamefont
  {Abouraddy}}, \bibinfo {author} {\bibfnamefont {Demetrios~N}\ \bibnamefont
  {Christodoulides}}, \ and\ \bibinfo {author} {\bibfnamefont {Bahaa~EA}\
  \bibnamefont {Saleh}},\ }\bibfield  {title} {\enquote {\bibinfo {title}
  {Sub-thermal to super-thermal light statistics from a disordered lattice via
  deterministic control of excitation symmetry},}\ }\href@noop {} {\bibfield
  {journal} {\bibinfo  {journal} {Optica}\ }\textbf {\bibinfo {volume} {3}},\
  \bibinfo {pages} {477--482} (\bibinfo {year} {2016})}\BibitemShut {NoStop}%
\bibitem [{\citenamefont {Bender}\ \emph {et~al.}(2018)\citenamefont {Bender},
  \citenamefont {Y{\i}lmaz}, \citenamefont {Bromberg},\ and\ \citenamefont
  {Cao}}]{bender2018customizing}%
  \BibitemOpen
  \bibfield  {author} {\bibinfo {author} {\bibfnamefont {Nicholas}\
  \bibnamefont {Bender}}, \bibinfo {author} {\bibfnamefont {Hasan}\
  \bibnamefont {Y{\i}lmaz}}, \bibinfo {author} {\bibfnamefont {Yaron}\
  \bibnamefont {Bromberg}}, \ and\ \bibinfo {author} {\bibfnamefont {Hui}\
  \bibnamefont {Cao}},\ }\bibfield  {title} {\enquote {\bibinfo {title}
  {Customizing speckle intensity statistics},}\ }\href@noop {} {\bibfield
  {journal} {\bibinfo  {journal} {Optica}\ }\textbf {\bibinfo {volume} {5}},\
  \bibinfo {pages} {595--600} (\bibinfo {year} {2018})}\BibitemShut {NoStop}%
\bibitem [{\citenamefont {Li}\ \emph {et~al.}(2021)\citenamefont {Li},
  \citenamefont {Nie}, \citenamefont {Yang}, \citenamefont {Liu}, \citenamefont
  {Liu}, \citenamefont {Dong}, \citenamefont {Zhao}, \citenamefont {Peng},
  \citenamefont {Zubairy},\ and\ \citenamefont {Scully}}]{li2021sub}%
  \BibitemOpen
  \bibfield  {author} {\bibinfo {author} {\bibfnamefont {Zheng}\ \bibnamefont
  {Li}}, \bibinfo {author} {\bibfnamefont {Xiaoyu}\ \bibnamefont {Nie}},
  \bibinfo {author} {\bibfnamefont {Fan}\ \bibnamefont {Yang}}, \bibinfo
  {author} {\bibfnamefont {Xiangpei}\ \bibnamefont {Liu}}, \bibinfo {author}
  {\bibfnamefont {Dongyu}\ \bibnamefont {Liu}}, \bibinfo {author}
  {\bibfnamefont {Xiaolong}\ \bibnamefont {Dong}}, \bibinfo {author}
  {\bibfnamefont {Xingchen}\ \bibnamefont {Zhao}}, \bibinfo {author}
  {\bibfnamefont {Tao}\ \bibnamefont {Peng}}, \bibinfo {author} {\bibfnamefont
  {M~Suhail}\ \bibnamefont {Zubairy}}, \ and\ \bibinfo {author} {\bibfnamefont
  {Marlan~O}\ \bibnamefont {Scully}},\ }\bibfield  {title} {\enquote {\bibinfo
  {title} {Sub-rayleigh second-order correlation imaging using spatially
  distributive colored noise speckle patterns},}\ }\href@noop {} {\bibfield
  {journal} {\bibinfo  {journal} {Opt. Express}\ }\textbf {\bibinfo {volume}
  {29}},\ \bibinfo {pages} {19621--19630} (\bibinfo {year} {2021})}\BibitemShut
  {NoStop}%
\bibitem [{\citenamefont {Nie}\ \emph {et~al.}(2021)\citenamefont {Nie},
  \citenamefont {Yang}, \citenamefont {Liu}, \citenamefont {Zhao},
  \citenamefont {Nessler}, \citenamefont {Peng}, \citenamefont {Zubairy},\ and\
  \citenamefont {Scully}}]{nie2021noise}%
  \BibitemOpen
  \bibfield  {author} {\bibinfo {author} {\bibfnamefont {Xiaoyu}\ \bibnamefont
  {Nie}}, \bibinfo {author} {\bibfnamefont {Fan}\ \bibnamefont {Yang}},
  \bibinfo {author} {\bibfnamefont {Xiangpei}\ \bibnamefont {Liu}}, \bibinfo
  {author} {\bibfnamefont {Xingchen}\ \bibnamefont {Zhao}}, \bibinfo {author}
  {\bibfnamefont {Reed}\ \bibnamefont {Nessler}}, \bibinfo {author}
  {\bibfnamefont {Tao}\ \bibnamefont {Peng}}, \bibinfo {author} {\bibfnamefont
  {M~Suhail}\ \bibnamefont {Zubairy}}, \ and\ \bibinfo {author} {\bibfnamefont
  {Marlan~O}\ \bibnamefont {Scully}},\ }\bibfield  {title} {\enquote {\bibinfo
  {title} {Noise-robust computational ghost imaging with pink noise speckle
  patterns},}\ }\href@noop {} {\bibfield  {journal} {\bibinfo  {journal}
  {Physical Review A}\ }\textbf {\bibinfo {volume} {104}},\ \bibinfo {pages}
  {013513} (\bibinfo {year} {2021})}\BibitemShut {NoStop}%
\bibitem [{\citenamefont {Luo}\ \emph {et~al.}(2018)\citenamefont {Luo},
  \citenamefont {Yin}, \citenamefont {Yin}, \citenamefont {Wu},\ and\
  \citenamefont {Guo}}]{luo2018orthonormalization}%
  \BibitemOpen
  \bibfield  {author} {\bibinfo {author} {\bibfnamefont {Bin}\ \bibnamefont
  {Luo}}, \bibinfo {author} {\bibfnamefont {Pengqi}\ \bibnamefont {Yin}},
  \bibinfo {author} {\bibfnamefont {Longfei}\ \bibnamefont {Yin}}, \bibinfo
  {author} {\bibfnamefont {Guohua}\ \bibnamefont {Wu}}, \ and\ \bibinfo
  {author} {\bibfnamefont {Hong}\ \bibnamefont {Guo}},\ }\bibfield  {title}
  {\enquote {\bibinfo {title} {Orthonormalization method in ghost imaging},}\
  }\href@noop {} {\bibfield  {journal} {\bibinfo  {journal} {Optics Express}\
  }\textbf {\bibinfo {volume} {26}},\ \bibinfo {pages} {23093--23106} (\bibinfo
  {year} {2018})}\BibitemShut {NoStop}%
\bibitem [{\citenamefont {Wang}\ and\ \citenamefont
  {Zhao}(2016)}]{wang2016fast}%
  \BibitemOpen
  \bibfield  {author} {\bibinfo {author} {\bibfnamefont {Le}~\bibnamefont
  {Wang}}\ and\ \bibinfo {author} {\bibfnamefont {Shengmei}\ \bibnamefont
  {Zhao}},\ }\bibfield  {title} {\enquote {\bibinfo {title} {Fast reconstructed
  and high-quality ghost imaging with fast walsh--hadamard transform},}\
  }\href@noop {} {\bibfield  {journal} {\bibinfo  {journal} {Photonics
  Research}\ }\textbf {\bibinfo {volume} {4}},\ \bibinfo {pages} {240--244}
  (\bibinfo {year} {2016})}\BibitemShut {NoStop}%
\bibitem [{\citenamefont {Zhang}\ \emph {et~al.}(2017)\citenamefont {Zhang},
  \citenamefont {Wang}, \citenamefont {Zheng},\ and\ \citenamefont
  {Zhong}}]{zhang2017hadamard}%
  \BibitemOpen
  \bibfield  {author} {\bibinfo {author} {\bibfnamefont {Zibang}\ \bibnamefont
  {Zhang}}, \bibinfo {author} {\bibfnamefont {Xueying}\ \bibnamefont {Wang}},
  \bibinfo {author} {\bibfnamefont {Guoan}\ \bibnamefont {Zheng}}, \ and\
  \bibinfo {author} {\bibfnamefont {Jingang}\ \bibnamefont {Zhong}},\
  }\bibfield  {title} {\enquote {\bibinfo {title} {Hadamard single-pixel
  imaging versus fourier single-pixel imaging},}\ }\href@noop {} {\bibfield
  {journal} {\bibinfo  {journal} {Optics Express}\ }\textbf {\bibinfo {volume}
  {25}},\ \bibinfo {pages} {19619--19639} (\bibinfo {year} {2017})}\BibitemShut
  {NoStop}%
\bibitem [{\citenamefont {Yu}(2019)}]{yu2019super}%
  \BibitemOpen
  \bibfield  {author} {\bibinfo {author} {\bibfnamefont {Wen-Kai}\ \bibnamefont
  {Yu}},\ }\bibfield  {title} {\enquote {\bibinfo {title} {Super sub-nyquist
  single-pixel imaging by means of cake-cutting hadamard basis sort},}\
  }\href@noop {} {\bibfield  {journal} {\bibinfo  {journal} {Sensors}\ }\textbf
  {\bibinfo {volume} {19}},\ \bibinfo {pages} {4122} (\bibinfo {year}
  {2019})}\BibitemShut {NoStop}%
\bibitem [{\citenamefont {Nie}\ \emph {et~al.}(2020)\citenamefont {Nie},
  \citenamefont {Zhao}, \citenamefont {Peng},\ and\ \citenamefont
  {Scully}}]{nie2020sub}%
  \BibitemOpen
  \bibfield  {author} {\bibinfo {author} {\bibfnamefont {Xiaoyu}\ \bibnamefont
  {Nie}}, \bibinfo {author} {\bibfnamefont {Xingchen}\ \bibnamefont {Zhao}},
  \bibinfo {author} {\bibfnamefont {Tao}\ \bibnamefont {Peng}}, \ and\ \bibinfo
  {author} {\bibfnamefont {Marlan~O}\ \bibnamefont {Scully}},\ }\bibfield
  {title} {\enquote {\bibinfo {title} {Sub-nyquist computational ghost imaging
  with orthonormalized colored noise pattern},}\ }\href@noop {} {\bibfield
  {journal} {\bibinfo  {journal} {arXiv preprint arXiv:2012.07250}\ } (\bibinfo
  {year} {2020})}\BibitemShut {NoStop}%
\bibitem [{Note1()}]{Note1}%
  \BibitemOpen
  \bibinfo {note} {The raw codes of Speckle-Net can be found on \protect \url
  {https://github.com/XJTU-TAMU-CGI/PatternDL}}\BibitemShut {NoStop}%
\bibitem [{\citenamefont {Nair}\ and\ \citenamefont
  {Hinton}(2010)}]{nair2010rectified}%
  \BibitemOpen
  \bibfield  {author} {\bibinfo {author} {\bibfnamefont {Vinod}\ \bibnamefont
  {Nair}}\ and\ \bibinfo {author} {\bibfnamefont {Geoffrey~E}\ \bibnamefont
  {Hinton}},\ }\bibfield  {title} {\enquote {\bibinfo {title} {Rectified linear
  units improve restricted boltzmann machines},}\ }in\ \href@noop {} {\emph
  {\bibinfo {booktitle} {Icml}}}\ (\bibinfo {year} {2010})\BibitemShut
  {NoStop}%
\bibitem [{\citenamefont {Ioffe}\ and\ \citenamefont
  {Szegedy}(2015)}]{ioffe2015batch}%
  \BibitemOpen
  \bibfield  {author} {\bibinfo {author} {\bibfnamefont {Sergey}\ \bibnamefont
  {Ioffe}}\ and\ \bibinfo {author} {\bibfnamefont {Christian}\ \bibnamefont
  {Szegedy}},\ }\bibfield  {title} {\enquote {\bibinfo {title} {Batch
  normalization: Accelerating deep network training by reducing internal
  covariate shift},}\ }in\ \href@noop {} {\emph {\bibinfo {booktitle}
  {International conference on machine learning}}}\ (\bibinfo {organization}
  {PMLR},\ \bibinfo {year} {2015})\ pp.\ \bibinfo {pages}
  {448--456}\BibitemShut {NoStop}%
\bibitem [{\citenamefont {Zhu}\ and\ \citenamefont {Bain}(2017)}]{BCNNpaper}%
  \BibitemOpen
  \bibfield  {author} {\bibinfo {author} {\bibfnamefont {Xinqi}\ \bibnamefont
  {Zhu}}\ and\ \bibinfo {author} {\bibfnamefont {Michael}\ \bibnamefont
  {Bain}},\ }\bibfield  {title} {\enquote {\bibinfo {title} {B-cnn: branch
  convolutional neural network for hierarchical classification},}\ }\href@noop
  {} {\bibfield  {journal} {\bibinfo  {journal} {arXiv preprint
  arXiv:1709.09890}\ } (\bibinfo {year} {2017})}\BibitemShut {NoStop}%
\bibitem [{\citenamefont {He}\ \emph {et~al.}(2016)\citenamefont {He},
  \citenamefont {Zhang}, \citenamefont {Ren},\ and\ \citenamefont
  {Jian}}]{2016ResNet}%
  \BibitemOpen
  \bibfield  {author} {\bibinfo {author} {\bibfnamefont {K.}~\bibnamefont
  {He}}, \bibinfo {author} {\bibfnamefont {X.}~\bibnamefont {Zhang}}, \bibinfo
  {author} {\bibfnamefont {S.}~\bibnamefont {Ren}}, \ and\ \bibinfo {author}
  {\bibfnamefont {S.}~\bibnamefont {Jian}},\ }\bibfield  {title} {\enquote
  {\bibinfo {title} {Identity mappings in deep residual networks},}\ }in\
  \href@noop {} {\emph {\bibinfo {booktitle} {European Conference on Computer
  Vision}}}\ (\bibinfo {year} {2016})\BibitemShut {NoStop}%
\bibitem [{Note2()}]{Note2}%
  \BibitemOpen
  \bibinfo {note} {For instance, if the size of each speckle pattern is
  $112\times 112$ pixels and the SR $\beta = 0.5\%$, the number of patterns is
  62. Then the size of parameters in the FC layer is around 9,000 TB, which is
  unrealistic for training.}\BibitemShut {Stop}%
\bibitem [{\citenamefont {Pittman}\ \emph {et~al.}(1995)\citenamefont
  {Pittman}, \citenamefont {Shih}, \citenamefont {Strekalov},\ and\
  \citenamefont {Sergienko}}]{pittman1995optical}%
  \BibitemOpen
  \bibfield  {author} {\bibinfo {author} {\bibfnamefont {Todd~B}\ \bibnamefont
  {Pittman}}, \bibinfo {author} {\bibfnamefont {YH}~\bibnamefont {Shih}},
  \bibinfo {author} {\bibfnamefont {DV}~\bibnamefont {Strekalov}}, \ and\
  \bibinfo {author} {\bibfnamefont {Alexander~V}\ \bibnamefont {Sergienko}},\
  }\bibfield  {title} {\enquote {\bibinfo {title} {Optical imaging by means of
  two-photon quantum entanglement},}\ }\href@noop {} {\bibfield  {journal}
  {\bibinfo  {journal} {Physical Review A}\ }\textbf {\bibinfo {volume} {52}},\
  \bibinfo {pages} {R3429} (\bibinfo {year} {1995})}\BibitemShut {NoStop}%
\bibitem [{\citenamefont {Bromberg}\ \emph {et~al.}(2009)\citenamefont
  {Bromberg}, \citenamefont {Katz},\ and\ \citenamefont
  {Silberberg}}]{bromberg2009ghost}%
  \BibitemOpen
  \bibfield  {author} {\bibinfo {author} {\bibfnamefont {Yaron}\ \bibnamefont
  {Bromberg}}, \bibinfo {author} {\bibfnamefont {Ori}\ \bibnamefont {Katz}}, \
  and\ \bibinfo {author} {\bibfnamefont {Yaron}\ \bibnamefont {Silberberg}},\
  }\bibfield  {title} {\enquote {\bibinfo {title} {Ghost imaging with a single
  detector},}\ }\href@noop {} {\bibfield  {journal} {\bibinfo  {journal}
  {Physical Review A}\ }\textbf {\bibinfo {volume} {79}},\ \bibinfo {pages}
  {053840} (\bibinfo {year} {2009})}\BibitemShut {NoStop}%
\bibitem [{\citenamefont {Zhang}\ \emph {et~al.}(2015)\citenamefont {Zhang},
  \citenamefont {Ma},\ and\ \citenamefont {Zhong}}]{zhang2015single}%
  \BibitemOpen
  \bibfield  {author} {\bibinfo {author} {\bibfnamefont {Zibang}\ \bibnamefont
  {Zhang}}, \bibinfo {author} {\bibfnamefont {Xiao}\ \bibnamefont {Ma}}, \ and\
  \bibinfo {author} {\bibfnamefont {Jingang}\ \bibnamefont {Zhong}},\
  }\bibfield  {title} {\enquote {\bibinfo {title} {Single-pixel imaging by
  means of fourier spectrum acquisition},}\ }\href@noop {} {\bibfield
  {journal} {\bibinfo  {journal} {Nature communications}\ }\textbf {\bibinfo
  {volume} {6}},\ \bibinfo {pages} {1--6} (\bibinfo {year} {2015})}\BibitemShut
  {NoStop}%
\bibitem [{\citenamefont {Katz}\ \emph {et~al.}(2009)\citenamefont {Katz},
  \citenamefont {Bromberg},\ and\ \citenamefont
  {Silberberg}}]{katz2009compressive}%
  \BibitemOpen
  \bibfield  {author} {\bibinfo {author} {\bibfnamefont {Ori}\ \bibnamefont
  {Katz}}, \bibinfo {author} {\bibfnamefont {Yaron}\ \bibnamefont {Bromberg}},
  \ and\ \bibinfo {author} {\bibfnamefont {Yaron}\ \bibnamefont {Silberberg}},\
  }\bibfield  {title} {\enquote {\bibinfo {title} {Compressive ghost
  imaging},}\ }\href@noop {} {\bibfield  {journal} {\bibinfo  {journal}
  {Applied Physics Letters}\ }\textbf {\bibinfo {volume} {95}},\ \bibinfo
  {pages} {131110} (\bibinfo {year} {2009})}\BibitemShut {NoStop}%
\bibitem [{\citenamefont {Katkovnik}\ and\ \citenamefont
  {Astola}(2012)}]{katkovnik2012compressive}%
  \BibitemOpen
  \bibfield  {author} {\bibinfo {author} {\bibfnamefont {Vladimir}\
  \bibnamefont {Katkovnik}}\ and\ \bibinfo {author} {\bibfnamefont {Jaakko}\
  \bibnamefont {Astola}},\ }\bibfield  {title} {\enquote {\bibinfo {title}
  {Compressive sensing computational ghost imaging},}\ }\href@noop {}
  {\bibfield  {journal} {\bibinfo  {journal} {JOSA A}\ }\textbf {\bibinfo
  {volume} {29}},\ \bibinfo {pages} {1556--1567} (\bibinfo {year}
  {2012})}\BibitemShut {NoStop}%
\bibitem [{\citenamefont {Lyu}\ \emph {et~al.}(2017)\citenamefont {Lyu},
  \citenamefont {Wang}, \citenamefont {Wang}, \citenamefont {Wang},
  \citenamefont {Li}, \citenamefont {Chen},\ and\ \citenamefont
  {Situ}}]{lyu2017deep}%
  \BibitemOpen
  \bibfield  {author} {\bibinfo {author} {\bibfnamefont {Meng}\ \bibnamefont
  {Lyu}}, \bibinfo {author} {\bibfnamefont {Wei}\ \bibnamefont {Wang}},
  \bibinfo {author} {\bibfnamefont {Hao}\ \bibnamefont {Wang}}, \bibinfo
  {author} {\bibfnamefont {Haichao}\ \bibnamefont {Wang}}, \bibinfo {author}
  {\bibfnamefont {Guowei}\ \bibnamefont {Li}}, \bibinfo {author} {\bibfnamefont
  {Ni}~\bibnamefont {Chen}}, \ and\ \bibinfo {author} {\bibfnamefont {Guohai}\
  \bibnamefont {Situ}},\ }\bibfield  {title} {\enquote {\bibinfo {title}
  {Deep-learning-based ghost imaging},}\ }\href@noop {} {\bibfield  {journal}
  {\bibinfo  {journal} {Scientific Reports}\ }\textbf {\bibinfo {volume} {7}},\
  \bibinfo {pages} {1--6} (\bibinfo {year} {2017})}\BibitemShut {NoStop}%
\bibitem [{\citenamefont {Shimobaba}\ \emph {et~al.}(2018)\citenamefont
  {Shimobaba}, \citenamefont {Endo}, \citenamefont {Nishitsuji}, \citenamefont
  {Takahashi}, \citenamefont {Nagahama}, \citenamefont {Hasegawa},
  \citenamefont {Sano}, \citenamefont {Hirayama}, \citenamefont {Kakue},
  \citenamefont {Shiraki} \emph {et~al.}}]{shimobaba2018computational}%
  \BibitemOpen
  \bibfield  {author} {\bibinfo {author} {\bibfnamefont {Tomoyoshi}\
  \bibnamefont {Shimobaba}}, \bibinfo {author} {\bibfnamefont {Yutaka}\
  \bibnamefont {Endo}}, \bibinfo {author} {\bibfnamefont {Takashi}\
  \bibnamefont {Nishitsuji}}, \bibinfo {author} {\bibfnamefont {Takayuki}\
  \bibnamefont {Takahashi}}, \bibinfo {author} {\bibfnamefont {Yuki}\
  \bibnamefont {Nagahama}}, \bibinfo {author} {\bibfnamefont {Satoki}\
  \bibnamefont {Hasegawa}}, \bibinfo {author} {\bibfnamefont {Marie}\
  \bibnamefont {Sano}}, \bibinfo {author} {\bibfnamefont {Ryuji}\ \bibnamefont
  {Hirayama}}, \bibinfo {author} {\bibfnamefont {Takashi}\ \bibnamefont
  {Kakue}}, \bibinfo {author} {\bibfnamefont {Atsushi}\ \bibnamefont
  {Shiraki}},  \emph {et~al.},\ }\bibfield  {title} {\enquote {\bibinfo {title}
  {Computational ghost imaging using deep learning},}\ }\href@noop {}
  {\bibfield  {journal} {\bibinfo  {journal} {Optics Communications}\ }\textbf
  {\bibinfo {volume} {413}},\ \bibinfo {pages} {147--151} (\bibinfo {year}
  {2018})}\BibitemShut {NoStop}%
\bibitem [{\citenamefont {Barbastathis}\ \emph {et~al.}(2019)\citenamefont
  {Barbastathis}, \citenamefont {Ozcan},\ and\ \citenamefont
  {Situ}}]{barbastathis2019use}%
  \BibitemOpen
  \bibfield  {author} {\bibinfo {author} {\bibfnamefont {George}\ \bibnamefont
  {Barbastathis}}, \bibinfo {author} {\bibfnamefont {Aydogan}\ \bibnamefont
  {Ozcan}}, \ and\ \bibinfo {author} {\bibfnamefont {Guohai}\ \bibnamefont
  {Situ}},\ }\bibfield  {title} {\enquote {\bibinfo {title} {On the use of deep
  learning for computational imaging},}\ }\href@noop {} {\bibfield  {journal}
  {\bibinfo  {journal} {Optica}\ }\textbf {\bibinfo {volume} {6}},\ \bibinfo
  {pages} {921--943} (\bibinfo {year} {2019})}\BibitemShut {NoStop}%
\bibitem [{\citenamefont {Wang}\ \emph {et~al.}(2019)\citenamefont {Wang},
  \citenamefont {Wang}, \citenamefont {Wang}, \citenamefont {Li},\ and\
  \citenamefont {Situ}}]{wang2019learning}%
  \BibitemOpen
  \bibfield  {author} {\bibinfo {author} {\bibfnamefont {Fei}\ \bibnamefont
  {Wang}}, \bibinfo {author} {\bibfnamefont {Hao}\ \bibnamefont {Wang}},
  \bibinfo {author} {\bibfnamefont {Haichao}\ \bibnamefont {Wang}}, \bibinfo
  {author} {\bibfnamefont {Guowei}\ \bibnamefont {Li}}, \ and\ \bibinfo
  {author} {\bibfnamefont {Guohai}\ \bibnamefont {Situ}},\ }\bibfield  {title}
  {\enquote {\bibinfo {title} {Learning from simulation: An end-to-end
  deep-learning approach for computational ghost imaging},}\ }\href@noop {}
  {\bibfield  {journal} {\bibinfo  {journal} {Optics Express}\ }\textbf
  {\bibinfo {volume} {27}},\ \bibinfo {pages} {25560--25572} (\bibinfo {year}
  {2019})}\BibitemShut {NoStop}%
\bibitem [{\citenamefont {Wu}\ \emph {et~al.}(2020)\citenamefont {Wu},
  \citenamefont {Wang}, \citenamefont {Zhao}, \citenamefont {Xiao},
  \citenamefont {Wang}, \citenamefont {Liang}, \citenamefont {Tian},
  \citenamefont {Cheng},\ and\ \citenamefont {Zhang}}]{wu2020sub}%
  \BibitemOpen
  \bibfield  {author} {\bibinfo {author} {\bibfnamefont {Heng}\ \bibnamefont
  {Wu}}, \bibinfo {author} {\bibfnamefont {Ruizhou}\ \bibnamefont {Wang}},
  \bibinfo {author} {\bibfnamefont {Genping}\ \bibnamefont {Zhao}}, \bibinfo
  {author} {\bibfnamefont {Huapan}\ \bibnamefont {Xiao}}, \bibinfo {author}
  {\bibfnamefont {Daodang}\ \bibnamefont {Wang}}, \bibinfo {author}
  {\bibfnamefont {Jian}\ \bibnamefont {Liang}}, \bibinfo {author}
  {\bibfnamefont {Xiaobo}\ \bibnamefont {Tian}}, \bibinfo {author}
  {\bibfnamefont {Lianglun}\ \bibnamefont {Cheng}}, \ and\ \bibinfo {author}
  {\bibfnamefont {Xianmin}\ \bibnamefont {Zhang}},\ }\bibfield  {title}
  {\enquote {\bibinfo {title} {Sub-nyquist computational ghost imaging with
  deep learning},}\ }\href@noop {} {\bibfield  {journal} {\bibinfo  {journal}
  {Optics Express}\ }\textbf {\bibinfo {volume} {28}},\ \bibinfo {pages}
  {3846--3853} (\bibinfo {year} {2020})}\BibitemShut {NoStop}%
\bibitem [{\citenamefont {Ruder}(2016)}]{ruder2016overview}%
  \BibitemOpen
  \bibfield  {author} {\bibinfo {author} {\bibfnamefont {Sebastian}\
  \bibnamefont {Ruder}},\ }\bibfield  {title} {\enquote {\bibinfo {title} {An
  overview of gradient descent optimization algorithms},}\ }\href@noop {}
  {\bibfield  {journal} {\bibinfo  {journal} {arXiv preprint arXiv:1609.04747}\
  } (\bibinfo {year} {2016})}\BibitemShut {NoStop}%
\bibitem [{\citenamefont {Ronneberger}\ \emph {et~al.}(2015)\citenamefont
  {Ronneberger}, \citenamefont {Fischer},\ and\ \citenamefont
  {Brox}}]{ronneberger2015u}%
  \BibitemOpen
  \bibfield  {author} {\bibinfo {author} {\bibfnamefont {Olaf}\ \bibnamefont
  {Ronneberger}}, \bibinfo {author} {\bibfnamefont {Philipp}\ \bibnamefont
  {Fischer}}, \ and\ \bibinfo {author} {\bibfnamefont {Thomas}\ \bibnamefont
  {Brox}},\ }\bibfield  {title} {\enquote {\bibinfo {title} {U-net:
  Convolutional networks for biomedical image segmentation},}\ }in\ \href@noop
  {} {\emph {\bibinfo {booktitle} {International Conference on Medical image
  computing and computer-assisted intervention}}}\ (\bibinfo {organization}
  {Springer},\ \bibinfo {year} {2015})\ pp.\ \bibinfo {pages}
  {234--241}\BibitemShut {NoStop}%
\bibitem [{\citenamefont {Connor}\ \emph {et~al.}(1994)\citenamefont {Connor},
  \citenamefont {Martin},\ and\ \citenamefont {Atlas}}]{connor1994recurrent}%
  \BibitemOpen
  \bibfield  {author} {\bibinfo {author} {\bibfnamefont {Jerome~T}\
  \bibnamefont {Connor}}, \bibinfo {author} {\bibfnamefont {R~Douglas}\
  \bibnamefont {Martin}}, \ and\ \bibinfo {author} {\bibfnamefont {Les~E}\
  \bibnamefont {Atlas}},\ }\bibfield  {title} {\enquote {\bibinfo {title}
  {Recurrent neural networks and robust time series prediction},}\ }\href@noop
  {} {\bibfield  {journal} {\bibinfo  {journal} {IEEE transactions on neural
  networks}\ }\textbf {\bibinfo {volume} {5}},\ \bibinfo {pages} {240--254}
  (\bibinfo {year} {1994})}\BibitemShut {NoStop}%
\bibitem [{\citenamefont {Jaderberg}\ \emph {et~al.}(2015)\citenamefont
  {Jaderberg}, \citenamefont {Simonyan}, \citenamefont {Zisserman} \emph
  {et~al.}}]{jaderberg2015spatial}%
  \BibitemOpen
  \bibfield  {author} {\bibinfo {author} {\bibfnamefont {Max}\ \bibnamefont
  {Jaderberg}}, \bibinfo {author} {\bibfnamefont {Karen}\ \bibnamefont
  {Simonyan}}, \bibinfo {author} {\bibfnamefont {Andrew}\ \bibnamefont
  {Zisserman}},  \emph {et~al.},\ }\bibfield  {title} {\enquote {\bibinfo
  {title} {Spatial transformer networks},}\ }\href@noop {} {\bibfield
  {journal} {\bibinfo  {journal} {Advances in neural information processing
  systems}\ }\textbf {\bibinfo {volume} {28}},\ \bibinfo {pages} {2017--2025}
  (\bibinfo {year} {2015})}\BibitemShut {NoStop}%
\bibitem [{\citenamefont {Vaswani}\ \emph {et~al.}(2017)\citenamefont
  {Vaswani}, \citenamefont {Shazeer}, \citenamefont {Parmar}, \citenamefont
  {Uszkoreit}, \citenamefont {Jones}, \citenamefont {Gomez}, \citenamefont
  {Kaiser},\ and\ \citenamefont {Polosukhin}}]{vaswani2017attention}%
  \BibitemOpen
  \bibfield  {author} {\bibinfo {author} {\bibfnamefont {Ashish}\ \bibnamefont
  {Vaswani}}, \bibinfo {author} {\bibfnamefont {Noam}\ \bibnamefont {Shazeer}},
  \bibinfo {author} {\bibfnamefont {Niki}\ \bibnamefont {Parmar}}, \bibinfo
  {author} {\bibfnamefont {Jakob}\ \bibnamefont {Uszkoreit}}, \bibinfo {author}
  {\bibfnamefont {Llion}\ \bibnamefont {Jones}}, \bibinfo {author}
  {\bibfnamefont {Aidan~N}\ \bibnamefont {Gomez}}, \bibinfo {author}
  {\bibfnamefont {{\L}ukasz}\ \bibnamefont {Kaiser}}, \ and\ \bibinfo {author}
  {\bibfnamefont {Illia}\ \bibnamefont {Polosukhin}},\ }\bibfield  {title}
  {\enquote {\bibinfo {title} {Attention is all you need},}\ }\href@noop {}
  {\bibfield  {journal} {\bibinfo  {journal} {Advances in neural information
  processing systems}\ ,\ \bibinfo {pages} {5998--6008}} (\bibinfo {year}
  {2017})}\BibitemShut {NoStop}%
\end{thebibliography}%

\ifarXiv


\section*{Supplementary material}

This document provides supplementary information to "Deep-learned speckle patterns for desired correlation and its application to ghost imaging". In the first section, we show the convergence of the loss function each round with different sampling ratios. In the second section, we discuss the features of speckle patterns resulting from the first and second training rounds and the CGI results using these patterns. In the third section, we discuss the effect of the initial pattern on the training process. In the fourth section, we examine the performance of Speckle-Net using a different loss function.

## SPECKLE-NET TRAINING

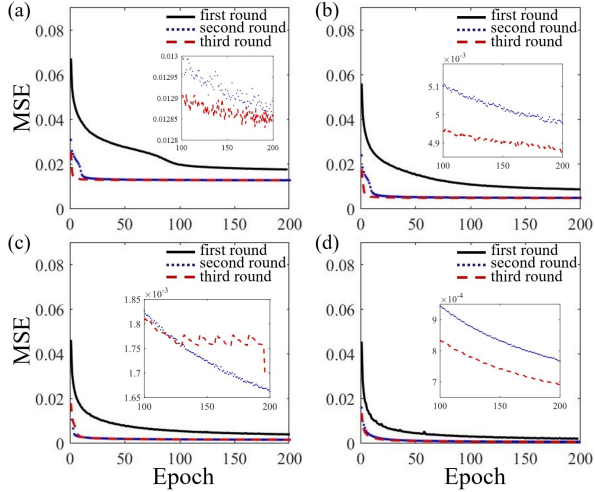

FIG. S1: (a), (b), (c), and (d) are the MSE as a function of epoch after each round of convolution with $\beta = 0.5\%$, 1%, 2%, and 5%, respectively. The first round of training is in solid black lines, the second round in blue dots, and the third round in red dotted lines.

At the end of every branch in training, the standard loss function, MSE, is adopted to give feedback to convolution layers in the neural network. Because of the abandonment of dropout layers, the over-fitting problem does not exist in our model. Therefore, we can let the training process operate as much as time allows or stop when the loss function tends to be convergence. In our experiment, we choose the training epoch equal to 200 in each branch based on the convergence speed of the loss function. In Fig. S1, we present the relationship between loss functions and training epoch according to $\beta = 0.5\%$, 1%, 2%, and 5% with each round of convolution. It is vividly shown that the first round convolution is always not enough. Comparing the three-round and two-round results, the loss function converges slightly better with the third-round results than the second-round results. Thus, we conclude that three rounds of training provide us with optimal results for all $\beta$ used in this study.

## SPECKLE PATTERNS WITH FIRST AND SECOND ROUND SPECKLE-NET TRAINING AND THE CGI EXPERIMENTAL RESULTS

### Speckle patterns with first and second round Speckle-Net training

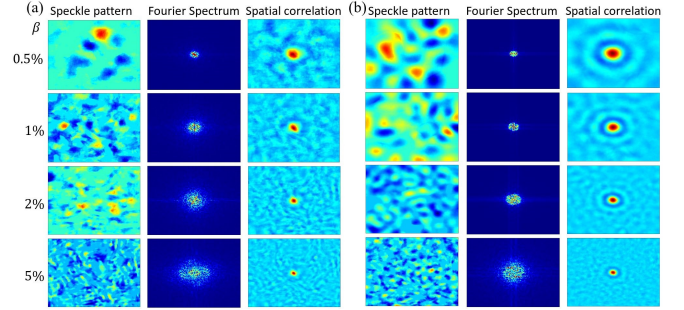

FIG. S2: (a) Typical first round speckle patterns, their Fourier spectra, and the spatial intensity correlation distributions from each $\beta$. (b) Typical second round speckle patterns, their Fourier spectra, and the spatial intensity correlation distributions from each $\beta$. From top to bottom, the sampling ratios are $\beta = 0.5\%, 1\%, 2\%$, and 5%.

The first and second round convoluted speckle patterns and the corresponding Fourier spectra, spatial intensity correlation distributions are shown in Fig. S2. It should be noted that the spatial correlation distribution still fluctuates, particularly in first round deep-learned speckle patterns, which indicates noise and blurriness in the imaging results.

From Fig. S2 we can see that the spatial intensity gradually turns to be lower under convolution with kernels. Moreover, the Fourier spectrum in Fig. S2 concentrate on low spatial frequency, and turn to be sharper on the edge of low frequency after two rounds training. Another typical feature of the convoluted patterns is that their spatial intensity fluctuation correlation distribution tends to be smoother when the round of training increases.

In addition, the bandwidth of correlation becomes larger, and it looks like the sinc function if we pay attention to its negative cross-correlation surrounding positive correlation at the center. These two features generated by DL provide second round convoluted speckle patterns with the capability of imaging with an extremely low sampling ratio.

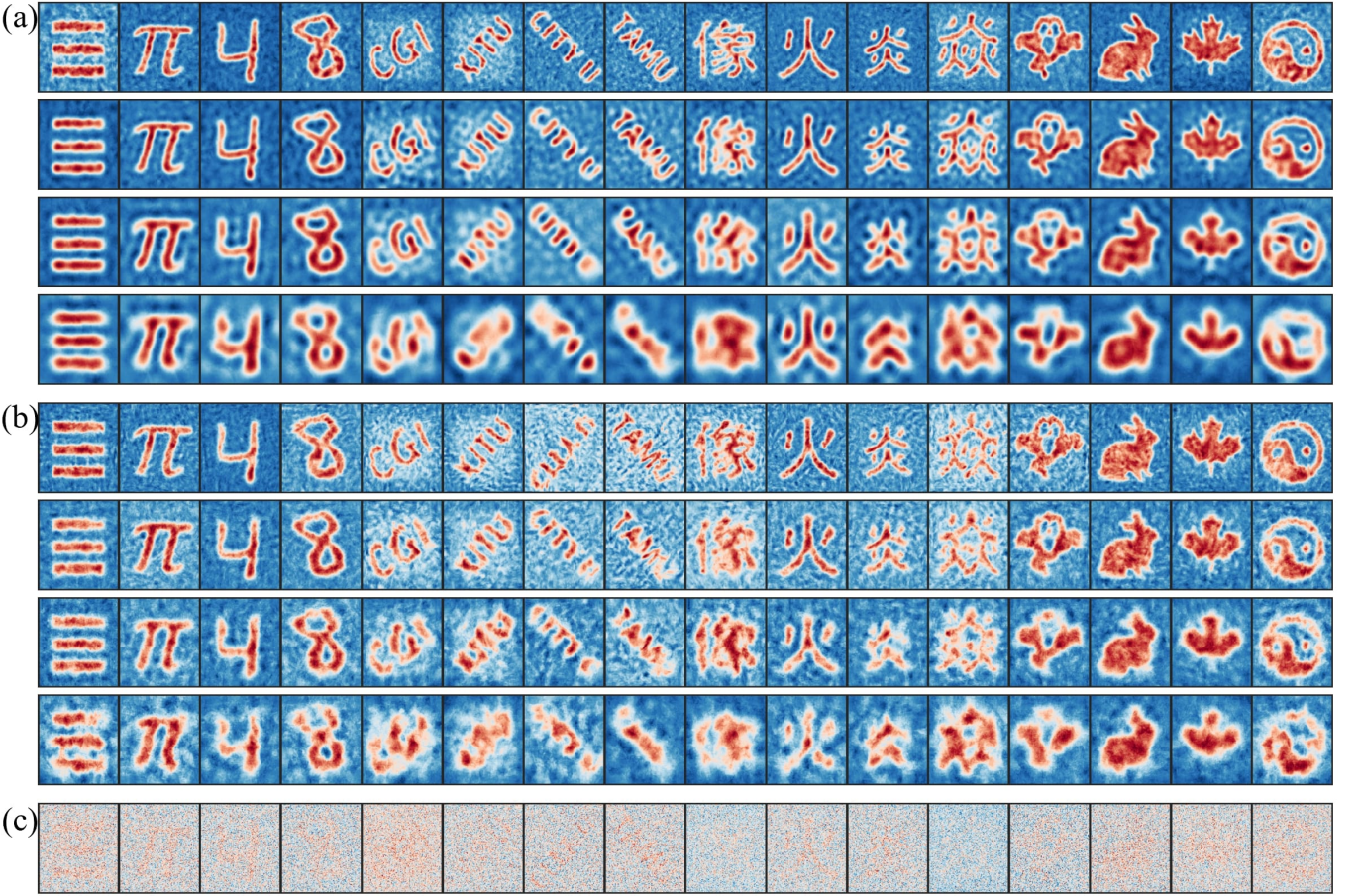

FIG. S3: (a) and (b): Experimental CGI results with the second round and first round deep-learned speckle patterns, respectively. The same 16 objects, including simple objects ('three lines', $\pi$, digits '4' and '8'), English letters ('CGI', 'XJTU', 'CITY U', and 'TAMU'), Chinese characters ('xiàng', 'hǔo', 'yán', and 'yàn', and pictures ('ghost', 'rabbit', 'leaf', and 'Tai Chi'), are used for the experiments. For each case, $\beta = 5\%$, $2\%$, $1\%$, and $0.5\%$, from top to bottom. (c): Experimental CGI results with commonly used white noise speckle patterns with $\beta = 5\%$.

**Experimental results using first and second round convoluted speckle patterns**

The first and second round trained speckle patterns are then used for the CGI experiments. CGI with conventional white noise speckle patterns are also measured for comparison. The results are shown in Fig. S3. Again, we note that the training database only contains handwriting digits, whereas the imaging targets are far more complicated and not included in the training database. In Fig S3 (c), we can conclude that at sampling rate $\beta = 5\%$, the conventional white noise speckle patterns cannot give any clear imaging results. In fact, usually white noise CGI cannot be distinguished until the sampling rate reach to 50%. However, our deep-learned speckle patterns with first round training already own ability to retrieve the object with high visibility as is shown in Fig. S3 (b), though the background fluctuation still exists. After two rounds training, we get much

clear results, as shown in Fig. S3 (a). The simple objects are successfully retrieved from $\beta = 0.5\%$, and the English letter, Chinese characters, and pictures can be retrieved at $\beta = 2\%$. Nevertheless, compared to results reconstructed from three-round speckle patterns shown in the main text, the results from second round speckle patterns have room for improvement.

**DEPENDENCE OF THE INITIAL PATTERN IN THE TRAINING PROCESS**

In the main text, we use a pink noise type speckle pattern as the initial pattern for the Speckle-Net training. Here we show that a typical white noise pattern can also be used for the training. Fig. S4 shows the relationship between loss functions and training epoch according to $\beta = 0.5\%$ and $1\%$, with each round of convolution. As compared to Fig. S1, we see that for both $\beta = 0.5\%$ and $1\%$, MSE converges to almost the same value after

three rounds of training. The only noticeable difference is in the first round. When a pink noise speckle pattern is used for the initial input, it has a lower MSE to start with and converges to a lower MSE value after the first round. However, the differences become negligible at the second and third rounds since the Kernels property becomes the dominant factor instead of the initial pattern itself.

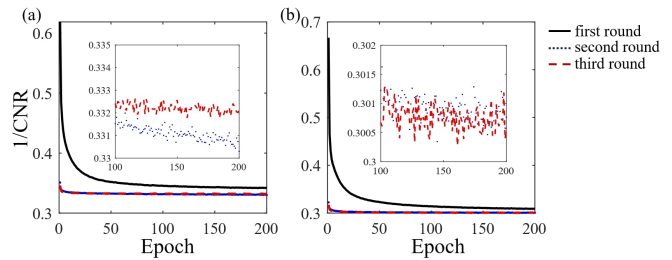

FIG. S6: 1/CNR as a function of epoch after each round of convolution with (a) $\beta = 0.5\%$, and (b) $\beta = 1\%$ respectively. The first round of training is in solid black lines, the second round in blue dots, and the third round in red dotted lines.

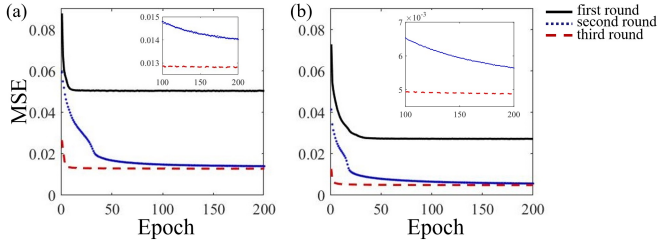

FIG. S4: (a), (b), and (c) are the loss function (MSE) as a function of epoch after each round of convolution with $\beta = 0.5\%$ and 1%, respectively. The first round of training is in solid black lines, the second round in blue dots, and the third round in dotted lines.

The features of the trained speckle pattern are also shown in Fig. S5. The first round trained speckle patterns are very similar to the white noise speckle patterns. The spectra are uniformly distributed except for the DC frequency region. The autocorrelation is peaked with a nonuniformly distributed background due to the insufficient ensemble average. The second and third round of trained speckle patterns are almost identical to those shown earlier, when the pink noise speckle pattern is used as the initial input.

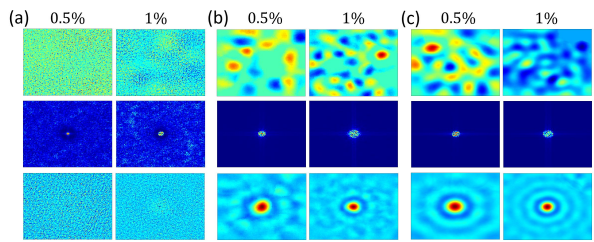

FIG. S5: first round (a), second round (b), and three-round(c) results when a typical white noise speckle pattern is used as the initial speckle pattern. From top to bottom lines: typical speckle patterns, their Fourier spectra, and the spatial intensity correlation distributions. The sampling rates corresponding to speckle patterns are $\beta = 0.5\%, 1\%$ labelled on the top.

## DIFFERENT LOSS FUNCTION IN SPECKLE-NET

Although MSE is the commonly used loss function for training, other quality indicators can also be used as the loss function in Speckle-Net. We here show that the contrast-to-noise ratio (CNR) function can also be used for the training process, and give very similar results as the MSE case. Different than SNR, the CNR tends to increase when the image quality increases. We therefore define the loss function as

$$\text{Loss} = 1/\text{CNR} = \frac{\sqrt{Var(G_{(o)}) + Var(G_{(b)})}}{\langle G_{(o)} \rangle - \langle G_{(b)} \rangle} \quad \text{(S1)}$$

Here $G$ represents pixels in the correlation results, $G_{(o)}$ is the object area, while $G_{(b)}$ is the background area. $\langle \cdots \rangle$ is the ensemble averaging over all the measurements. $Var(\cdots)$ is the variance of its arguments.

We also plot the first and second round convoluted speckle patterns and the corresponding Fourier spectra, spatial intensity correlation distributions are shown in Fig. S7. There is no obvious different as compared to the results shown in Fig. S2.

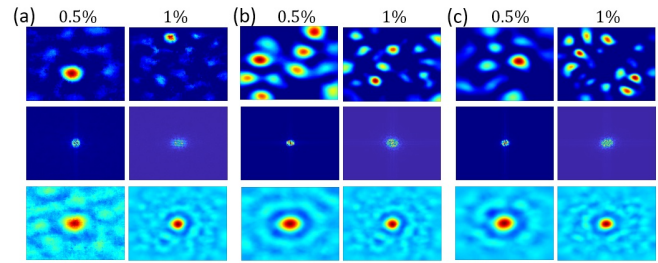

FIG. S7: first round (a), second round(b), and three-round(c) results when 1/CNR is used as the loss function. From top to bottom lines: typical speckle patterns, their Fourier spectra, and the spatial intensity correlation distributions. The sampling rates corresponding to speckle patterns are $\beta = 0.5\%, 1\%$ labelled on the top.


\fi

\end{document}